\UseRawInputEncoding
\documentclass[aps,reprint,floatfix,superscriptaddress]{revtex4-2}
\usepackage[english]{babel}
\usepackage[utf8]{inputenc}                             
\usepackage[T1]{fontenc}
\usepackage{bbold}
\usepackage{epsfig}
\usepackage{amsmath}
\usepackage{hyphenat}
\usepackage{units}
\usepackage[colorlinks]{hyperref}
\usepackage{placeins}
\usepackage{enumerate}
\usepackage{bbold}
\usepackage{caption}	
\usepackage[colorinlistoftodos]{todonotes}	
\usepackage{xcolor}	
\usepackage{color}

\newcommand{\abs}[1]{\left|{#1}\right|}

\begin{document}
	
	\title{Hydrodynamics of charged two-dimensional Dirac systems I: \\ thermo-electric transport}
	\author{Kitinan Pongsangangan}
	\affiliation{Institute for Theoretical Physics and Center for Extreme Matter and Emergent Phenomena,
		Utrecht University, Princetonplein 5, 3584 CC Utrecht, The Netherlands}
	\author{T. Ludwig}
	\affiliation{Institute for Theoretical Physics and Center for Extreme Matter and Emergent Phenomena,
		Utrecht University, Princetonplein 5, 3584 CC Utrecht, The Netherlands}
	\author{H.T.C. Stoof}
	\affiliation{Institute for Theoretical Physics and Center for Extreme Matter and Emergent Phenomena,
		Utrecht University, Princetonplein 5, 3584 CC Utrecht, The Netherlands}
	\author{Lars Fritz}
	\affiliation{Institute for Theoretical Physics and Center for Extreme Matter and Emergent Phenomena,
		Utrecht University, Princetonplein 5, 3584 CC Utrecht, The Netherlands}

	\begin{abstract}
	In this paper we study thermo-electric transport in interacting two-dimensional Dirac-type systems using a phenomenological Boltzmann approach. We consider a setup that can accommodate electrons, holes, and collective modes. In the first part of the paper we consider the electron-hole hydrodynamics, a model that is popular in the context of graphene, and its transport properties. In a second part, we propose a novel type of hydrodynamics. In that setup, the `fluid' consists of electrons, holes, and plasmons. We study its transport properties, especially the thermo-electric behavior. The results of this part can also be adapted to the study of a fluid consisting of electrons and phonons. This paper is accompanied by a technical paper in which we give a detailed derivation of the Boltzmann equations and the encoded conservation laws.
	\end{abstract}
	\maketitle
	
	\section{Introduction}
Hydrodynamic behavior of interacting electrons in solid-state systems has received a renewal of interest in recent years. This came from theorists and experimentalists alike, but was mostly driven by an increasing number of experimental observations on a broad range of materials~\cite{Molenkamp1995,Gooth2018,Jaoui2021}. The prime representative in terms of materials has been graphene and to a lesser extent bilayer graphene~\cite{Nam2017}. Although the physical conditions required for observing hydrodynamic behavior of electrons is demanding, the advancement in synthesizing nearly clean graphene sheets encapsulated in boron nitride~\cite{Dean2010} allowed the community to
explore several electron hydrodynamic phenomena over
the past few years~\cite{Crossno2016,Ghahari2016,Bandurin2016,Bandurin2018,Gallagher2019,Sulpizio2019}. Besides being superclean with very little disorder scattering, the concentration
of phonons is also suppressed in those devices. Consequently, the usual momentum relaxation of electrons by means of collisions with impurities and phonons becomes less frequent. As a result, electron-electron collisions which conserve total charge density, total momentum density, and total energy density can become the dominant relaxational process~\cite{Landau1965,KadanoffBaym1962,Fritz2008a,Fritz2008b,Kashuba2008}. If that is the case, the dynamics of the electrons can be described by the hydrodynamic transport theory of a typical fluid~\cite{LandauLifshitz1987,Gurzhi1963,Lucas2018,Narozhny2015,Narozhny2019}.

There is a different set of systems, that has been discussed recently. It uses phonons to its advantage: instead of destroying the hydrodynamic behavior, it is proposed that, when the electron-phonon collision is the dominant scattering process and phonons cannot relax momentum in the lattice or from higher-order scattering, electrons and phonons transfer their momentum between each other and form a single strongly coupled fluid that can again be described by a hydrodynamic theory. This electron-phonon hydrodynamics reveals unique transport properties which are different from the electron hydrodynamics; for example, in the temperature dependence of thermo-electric conductivities~\cite{gurevich1989,Levchenko2020,LucasHuang2021}. 

Collective oscillations of particles about their equilibrium position, like phonons, are ubiquitous phenomena in solid-state systems~\cite{Pines1999}. In addition to phonons that arise from lattice oscillations, there are also emergent collective excitations in interacting-electron systems; for example, plasmons and/or magnons. This naturally leads to a question: can exotic hydrodynamics arise also from a combined system of electrons and their collective modes? These excitations are bosonic in nature. So, in thermal equilibrium, they obey the Bose-Einstein distribution. Furthermore, they have no obvious innate relaxation mechanism; unlike phonons, which have the  lattice. The interplay between the collective modes and the electrons may give rise to rich effects in transport phenomena. One such example is the interplay between magnons and itinerant electrons in metallic magnetic heterostructures that has been studied in the context of spintronics~\cite{Been2022}. Here, we focus on plasmons. 
 
The problem of interactions between electrons and plasmons was put forth by a series of seminal works by Bohm and Pines in the 1960s~\cite{Pines&Bohm1952,Bohm&Pines1953,Pines&Schrieffer1962,WyldPines1962,Pines1953}. Recently, some of us revisited this problem; we presented and solved the kinetic theory for a coupled system of electrons and plasmons in two-dimensional interacting Dirac electrons. In particular, it was shown that in heat-transport probes, plasmons make a direct contribution that can potentially be of the same order of magnitude as the electronic contributions~\cite{Kitinan2020}. 

In this paper, we theoretically investigate the full thermo-electric transport properties of an interacting system of electrons and a collective mode that are coupled via perfect drag in the hydrodynamic limit. We use an approach based on a Boltzmann transport equation with a relaxation-time approximation. We treat electrons and bosons on an equal footing and discuss the hydrodynamic behavior of the combined system. While being particularly interested in the hydrodynamics of the coupled system of electrons and plasmons, we develop a theory that can be applied straighforwardly to any hybrid system of electrons, holes, and bosons.

The organization of the paper is as follows. In Sec.~\ref{sec:formalism} we introduce the general framework that allows us to treat both: an electron-hole system; and a system of electrons and holes that are drag-coupled to a bosonic mode. The approach is based on a Boltzmann equation and we discuss all the conservation laws of the underlying system. We also introduce a suitable relaxation-time approximation which allows to `fake' a full numerical solution of the Boltzmann equations by virtue of respecting all the underlying conservation laws. In Sec.~\ref{sec:electronholeplasma} we discuss the electron-hole plasma. This section is particularly important for the study of weakly interacting graphene close to the Dirac point. We study its thermo-electric response in Sec.~\ref{subsec:thermoelectricI} with a special eye on the Wiedemann-Franz ratio~\cite{WiedemannFranz1853}. We proceed to study the collective excitation of the charged electron-hole plasma taking into account the coupling to the classical Coulomb interaction in Sec.~\ref{subsec:plasmons}. In particular, we recover the plasmon dispersion that can be obtained from a more formal approach, the random-phase approximation (RPA). There is a second part of the paper, in which we discuss a novel scenario for hydrodynamics, which consists of electrons, holes, and plasmons. This scenario is particularly relevant to graphene but in principle also extends to other superclean two-dimensional electronic systems. We first discuss the theoretical setup in Sec.~\ref{subsec:setup} and then proceed to study the thermo-electric response of the drag-coupled fluid in Sec.~\ref{subsec:thermoelectricII}. This part constitutes the main result of this paper. It can be easily adapted; for example, to describe the thermo-electric response of a strongly drag-coupled electron-phonon liquid, which has recently been discussed in the literature~\cite{gurevich1989,Levchenko2020,LucasHuang2021}. We finish the paper with a conclusion and an outlook in Sec.~\ref{sec:conclusion}.

\section{Formalism}\label{sec:formalism}

	\begin{figure}
	\includegraphics[width=0.5\textwidth]{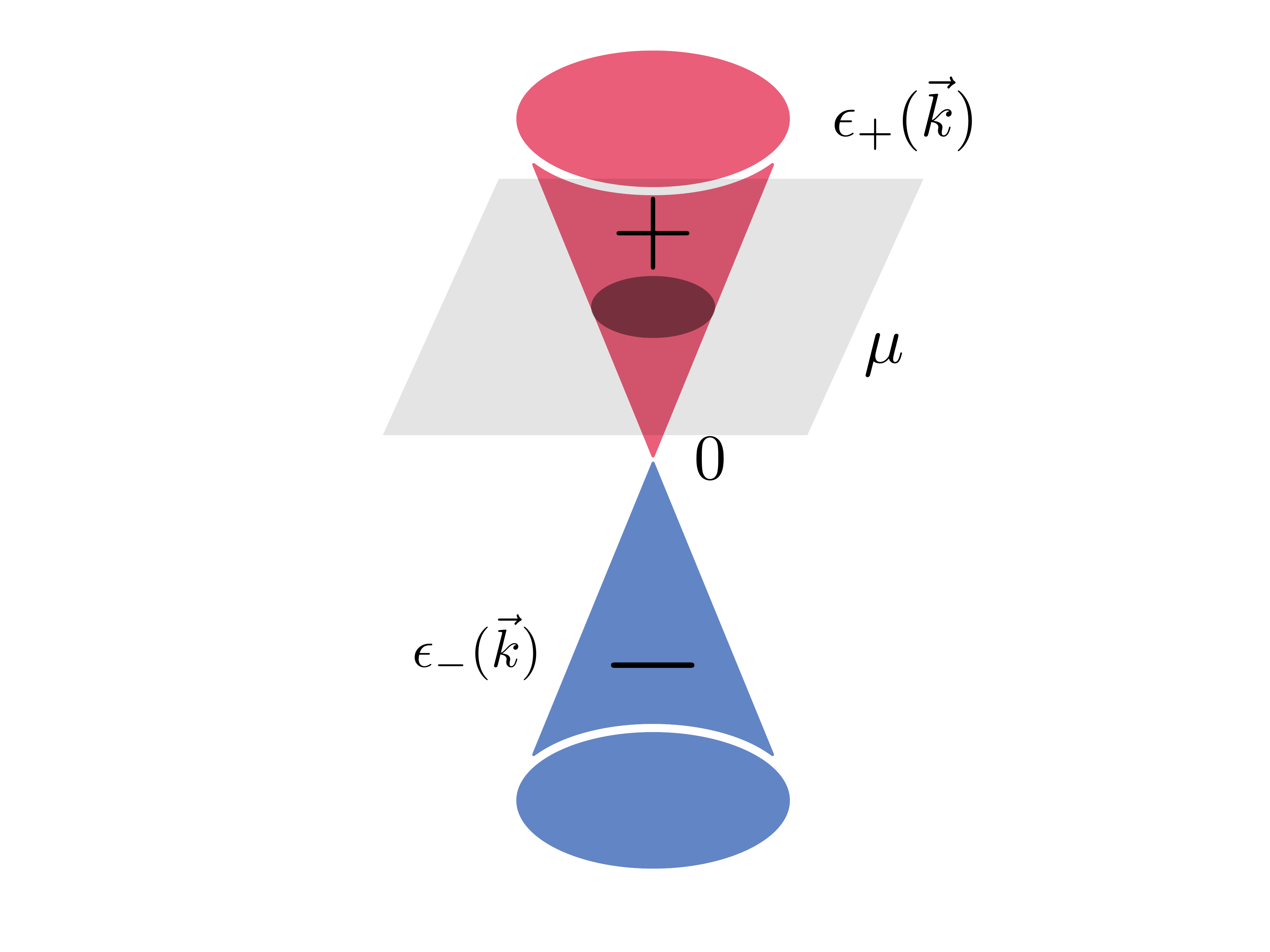}
	\caption{Generic two band system. We assume that there is an electron band ($+$) and a hole band ($-$) that touch in isolated points in the Brilllouin zone. Additionally, in our analysis we allow for a chemical potential $\mu$ that can be tuned into either band.}\label{fig:dispersion}
\end{figure}
Throughout this work, we use a phenomenological version of the Boltzmann-equation approach. We use it to calculate the thermo-electric response within two scenarios: (i) a two-band system with electrons and holes interacting with each other and (ii) an interacting system of electrons and holes in combination with a collective mode.

We consider an electronic two-band system consisting of conduction and valence electrons with their bands touching at isolated points in the Brillouin zone, as depicted schematically in Fig. \ref{fig:dispersion}. We assume that the valence band is completely filled and the conduction band is completely empty when the system is undoped. In that case, the Fermi energy is located at the touching point. Hereafter, we refer to this as the charge neutrality point and the chemical potential at this point is set to zero for convenience, {\it i.e.}, $\mu=0$. Furthermore, we assume particle-hole symmetry and rotational invariance around this point, {\it i.e.}, $\epsilon_{+}(\vec{k})=-\epsilon_{-}(\vec{k})$. An excitation of the type $\epsilon_+(\vec{k})$ is henceforth referred to as `electron' whereas $\epsilon_-(\vec{k})$ is called a `hole', as will become more clear below. 
 
 In addition, we have a bosonic mode associated with collective oscillations with a dispersion relation $\omega(\vec{k})$. The collective mode can be a lattice excitation like a phonon or an excitation of the underlying electronic system, for instance a plasmon and/or a spin wave. We assume that both the electron and boson are long-lived; meaning their distribution functions are well-defined. The central objects in the Boltzmann theory are these distribution functions. In equilibrium, the distribution functions for the original electrons reduce to the Fermi-Dirac distribution 
\begin{equation}
	f^0(\epsilon_{\pm}(\vec{k}))= \frac{1}{e^{\frac{(\epsilon_{\pm}(\vec{k})-\mu)}{k_B T}}+1},
\end{equation}
where $k_B$ is the Boltzmann constant and $\mu$ is the chemical potential. The boson distribution function assumes the form of the Planck or Bose-Einstein distribution 
\begin{equation}
	b^0(\omega(\vec{k}))=\frac{1}{e^{ \frac{\omega(\vec{k})}{k_B T}}-1}
\end{equation}
in equilibrium. A subtle point for the analysis of our model is that the two-band system is unbounded from below. This implies that the number of particles in the filled band is infinite, which would lead to a divergence in any calculations of physical observables. In order to circumvent this difficulty, we subtract `$1$' from the hole distribution function which amounts to subtracting the filled lower band and ensures that we can refrain from using a cutoff.   
We introduce the notation $f_{+}(\vec{k})=f (\epsilon_+(\vec{k}))$ for the electrons, $f_{-}(\vec{k})=f(\epsilon_-(\vec{k}))-1$ for the holes. In turn, in thermal equilibrium, we find
\begin{equation}
	f^0_\pm(\vec{k})=\pm \frac{1}{e^{\frac{\pm(\epsilon_{\pm}(\vec{k})-\mu)}{k_B T}}+1}\;.
\end{equation}
In general, the distribution functions,  denoted by $f_\pm(\vec{x},\vec{k},t)$ and $b(\vec{x},\vec{k},t)$, vary in space and time. They give the probability density that a particle is found in a phase-space volume element around the phase-space point $(\vec{x}, \vec{k})$ at time $t$. The time evolution of the distribution functions  is governed by a coupled system of Boltzmann equations,
\begin{widetext}
\begin{eqnarray}\label{eq:coupledBoltzmannfull}
\partial_t f_+(\vec{x},\vec{k},t)+ \partial_{\vec{k}} \epsilon_+(\vec{x},\vec{k}, t) \cdot \partial_{\vec{x}} f_+(\vec{x},\vec{k},t) - \partial_{\vec{x}} \epsilon_+ (\vec{x},\vec{k}, t)\cdot\partial_{\vec{k}} f_+(\vec{x},\vec{k},t)&=&\mathcal{C}^{ee}_++\mathcal{C}^{eh}_+ + \mathcal{C}^{b}_++\mathcal{C}^{\rm{dis}}_+ \;, \nonumber \\
\partial_t f_-(\vec{x},\vec{k},t)+ \partial_{\vec{k}} \epsilon_-(\vec{x},\vec{k}, t) \cdot \partial_{\vec{x}} f_-(\vec{x},\vec{k},t) -\partial_{\vec{x}} \epsilon_-  (\vec{x},\vec{k}, t)\cdot\partial_{\vec{k}} f_-(\vec{x},\vec{k},t)&=& \mathcal{C}^{hh}_-+\mathcal{C}^{he}_- + \mathcal{C}^{b}_-+\mathcal{C}^{\rm{dis}}_- \;, \nonumber \\ \partial_t b(\vec{x},\vec{k},t)+ \partial_{\vec{k}} \omega(\vec{x},\vec{k}, t) \cdot  \partial_{\vec{x}} b(\vec{x},\vec{k},t) -\partial_{\vec{x}} \omega (\vec{x},\vec{k}, t)\cdot\partial_{\vec{k}} b(\vec{x},\vec{k},t)&=&\mathcal{C}^{ee}_b+\mathcal{C}^{hh}_b +\mathcal{C}^{eh}_b +\mathcal{C}_b^{b}\;.
\end{eqnarray}
\end{widetext}
We use $\hbar=k_B=1$ here and throughout the remainder of the paper unless stated otherwise. The left-hand sides are the so-called streaming terms resulting from forces, inhomogeneities, and temporal changes. We assume that the energies of the particles can be space and time dependent. This allows us to account for external applied forces (for example, an external electric field) as well as for internal forces (for example, the Hartree potential between the electrons themselves).  It is worthwhile pointing out that in a two-band system, we also encounter Berry phase terms that we disregard here. They mostly matter for transverse response coefficients which we do not consider here. The right-hand sides in Eq.~\eqref{eq:coupledBoltzmannfull} describe the collisions, encoded in the collision integrals $\mathcal{C}_+$, $\mathcal{C}_-$, and $\mathcal{C}_b$, that enable the system to relax towards local thermal equilibrium. 

The collision terms have very different physical origin. There are inelastic processes between electrons, $\mathcal{C}^{ee}_+$,  between holes, $\mathcal{C}^{hh}_-$, and those mixing electrons and holes, $\mathcal{C}^{eh}_+$ and $\mathcal{C}^{he}_-$. Furthermore, electrons and holes undergo elastic disorder scattering, $\mathcal{C}^{\rm{dis}}_+$ and $\mathcal{C}^{\rm{dis}}_-$. Additionally, electrons and holes interact with bosons, $ \mathcal{C}^{b}_+$ and $ \mathcal{C}^{b}_-$ and vice versa according to $\mathcal{C}^{ee}_b$, $\mathcal{C}^{hh}_b$, and $\mathcal{C}^{eh}_b$. The collision integrals can be derived, for example, by using Fermi's Golden Rule. Instead of specifying a concrete scattering integral, our discussion is centred around the conservation laws of the system. There are a number of important assumptions here: 
\begin{enumerate}
\item To lowest order, disorder couples to electrons and holes only. It breaks translational symmetry and thereby momentum conservation but respects particle number and energy conservation.
\item Interactions between like particles, electron-electron or hole-hole respect total charge, individual particle number, momentum, and energy conservation.
\item Interactions between electrons and holes can transfer (`drag') momentum and energy from the electron sector to the hole sector and vice versa. Meanwhile, the total momentum and total energy within the combined electron-hole system are conserved. Equivalently, the number of electrons and holes, individually, is not necessarily conserved, only the total charge is.
\item The interaction between electrons, holes and bosons can transfer momentum and energy between all the three sectors; again representing drag.
\end{enumerate}

These statements relate to collisional invariants of the scattering integrals~\cite{Lifshitz1981}. These collisional invariants are obtained via momentum integrations over the collision integral and the corresponding conserved quantity. The corresponding integrals are of the type $\int \frac{d^dk}{(2\pi)^d}(1,\vec{k},\epsilon_+(\vec{k}))\mathcal{C}_+^{ee}=(N_+^{ee},K_+^{ee},E_+^{ee})$. The first integral refers to particle number ($N$), the second to momentum ($K$), and the last one to energy ($E$). Table \ref{tab1} shows the most generic form of these collisional invariants. We use this in Sec.~{\ref{sec:relaxationtime}} to construct a relaxation-time approximation that respects these properties.

Conservation laws can be derived by also integrating the left-hand sides of the Boltzmann equation over the collisional invariants, which are 1, $\vec{k}$, and the corresponding dispersion relation. The terms that appear from the streaming terms are: densities, generalized currents, and forces and heating. Instead of considering these quantities only for electrons, holes, and bosons, it is useful to also consider the total charge and the charge imbalance, as they play a role later. All these quantities are summarized in Table \ref{tab2} for densities and forces, and Table \ref{tab3} for currents for generic dispersions. 

\begin{widetext}
	\begin{center}
		\begin{table}[thb!]
			\tabcolsep7.5pt
			\caption{Collisional invariants}
			\label{tab1}
			\begin{center}
				\begin{tabular}{@{}l|c|c|clclc@{}}
					\hline
					&Particle number &Momentum  & Energy   \\
					\hline
					Electron &$N^{ee}_+,N^{\rm{dis}}_+=0 $& $K^{ee}_+=0$ & $E^{ee}_+, E_+^{\rm{dis}}=0$\\
					&$N_+^{b},N_+^{eh}\neq 0$& $K_+^{ee},K_+^{eh},K_+^{b}, K_+^{\rm{dis}} \neq0$ & $E_+^{ee},E_+^{eh},E_+^{b}\neq0$  \\
					Hole  &$N^{hh}_-,N^{\rm{dis}}_-=0$ &$K^{hh}_-=0$ & $E^{hh}_-, E_-^{\rm{dis}}=0$  \\ &$N_-^{b},N_-^{he}\neq 0$& $K_-^{hh},K_-^{he},K_-^{b},K_-^{\rm{dis}}\neq0$ & $E_-^{hh},E_-^{he},E_-^{b}\neq0$ \\
					Boson &$ N_b^{ee},N_b^{hh},N_b^{eh},N_b^{b} \neq 0$ &$K_b^{ee},K_b^{hh},K_b^{eh},K_b^{b} \neq 0$ & $E_b^{ee},E_b^{hh},E_b^{eh},E_b^{b} \neq 0$ \\
					\hline
		\end{tabular}
		\end{center}
	\end{table}
\end{center}
\end{widetext}

\subsubsection{Conservation laws}

The Boltzmann equations provide a convenient starting point for the derivation of the equations of hydrodynamics. As mentioned before, the basis of the whole discussion are the conservation laws. They do not only constitute a set of hydrodynamic equations, they also allow for an identification of physical quantities such as charge-current and energy-current densities. The conservation laws are straightforward to derive from the contents of Tables~\ref{tab1}-\ref{tab3}. The conservation laws are independent of the dispersion and can be written in full generality. This is not true for the Navier-Stokes equation. It has to be derived on a case-by-case basis and its concrete form depends on the dispersion relation of the underlying system, as explained below.

By integrating the Boltzmann equations for electrons and holes over all momenta $\vec{k}$, multiplying by $-e$, and adding them up, we obtain the conservation law of charge:
\begin{equation}
	\partial_t n_c + \partial_{\vec{x}} \cdot \vec{j}_c = 0\;.
\end{equation}
Next, we multiply all three Boltzmann equations of Eq.~\eqref{eq:coupledBoltzmannfull} with their respective energy, integrate the resulting equation over all states, and then add them together. In the end, we obtain the conservation law of the total energy, 
\begin{equation}
	\partial_t n_{\rm{tot}}^\epsilon+ \partial_{\vec{x}}\cdot\vec{j}_{\rm{tot}}^\epsilon = 0\;.
\end{equation}
Next, by multiplying Eq.~\eqref{eq:coupledBoltzmannfull} with momentum $\vec{k}$, integrating, and then summing all up, we obtain the conservation law of momentum according to 
\begin{eqnarray}
	&&	\partial_{t} \vec{n}_{\rm{tot}}^{\vec{k}} + {\partial_{\vec{x}} \cdot \vec{\vec{\Pi}}}_{\rm{tot}} =\vec{F}_c+\vec{F}_b+\vec{K}^{\rm{dis}}_++\vec{K}^{\rm{dis}}_-. 
\end{eqnarray}
The total momentum density is not locally conserved but changes by the internal electric forces on the right-hand side of the equation. However, it is required that the internal forces cancel each other when they are integrated over the whole space ${\vec{x}}$. This implies that
\begin{equation}
	\int_{\vec{x}} \left( \vec{F}_c+\vec{F}_b \right)= 0\;.
\end{equation}
Furthermore, the collision of electrons and holes with impurities relaxes the total momentum density towards equilibrium. This effect is encoded in the collision terms $\vec{K}^{\rm{dis}}_++\vec{K}^{\rm{dis}}_-$.
The definition of densities, forces, and currents are summarized in Tables \ref{tab2} and \ref{tab3}.
\begin{widetext}
\begin{center}
\begin{table}[hbt!]
		\tabcolsep7.5pt
		\caption{Densities and Forces}
		\label{tab2}
		\begin{center}
			\begin{tabular}{@{}l|c|c|c|clclc@{}}
				\hline
				&Density &Momentum density & Energy Density & 'Force' \\
				\hline
				Electron &$n_+= \int_{\vec{k}} f_+$& $\vec{n}^{\vec{k}}_+= \int_{\vec{k}} \vec{k} \;f_+$ & $n^\epsilon_+=\int_{\vec{k}} \epsilon_+ (\vec{x},\vec{k}, t) f_+$ & $\vec{F}_{+}=-\int_{\vec{k}}\partial_{\vec{x}}\epsilon_+(\vec{x},\vec{k}, t)\;f_{+}$ \\
				Hole  &$n_-= \int_{\vec{k}} f_-$ &$\vec{n}^{\vec{k}}_-= \int_{\vec{k}} \vec{k} \;f_-$ & $n^\epsilon_-=\int_{\vec{k}} \epsilon_- (\vec{x},\vec{k},  t) f_-$ &$\vec{F}_{-}=-\int_{\vec{k}}\partial_{\vec{x}}\epsilon_-(\vec{x},\vec{k},  t)\;f_{-}$ \\
				Charge &$n_c=-e\left(n_++n_-\right)$ &$\vec{n}^{\vec{k}}_c=\vec{n}^{\vec{k}}_++\vec{n}^{\vec{k}}_- $ & $n^\epsilon_c=n^\epsilon_++n^\epsilon_-$ &$\vec{F}_{c}=\vec{F}_{+}+\vec{F}_{-}$   \\
				Imbalance &$n_{\rm{imb}}=-e\left(n_+-n_-\right) $& $\vec{n}^{\vec{k}}_{\rm{imb}}=\vec{n}^{\vec{k}}_+-\vec{n}^{\vec{k}}_- $ &$n^\epsilon_{\rm{imb}}=n^\epsilon_+-n^\epsilon_-$ &$\vec{F}_{\rm{imb}}=\vec{F}_{+}-\vec{F}_{-}$  \\
				Boson &$n_b= \int_{\vec{k}} b$& $\vec{n}^{\vec{k}}_b= \int_{\vec{k}} \vec{k} \;b $ & $n^\epsilon_b=\int_{\vec{k}} \omega (\vec{x},\vec{k},  t) b$ &$\vec{F}_{b}=-\int_{\vec{k}}\partial_{\vec{x}}\omega(\vec{x},\vec{k},  t)\;b$  \\
				Total & - & $\vec{n}^{\vec{k}}_{\rm{tot}}=\vec{n}^{\vec{k}}_c+\vec{n}^{\vec{k}}_b$ &  $n^\epsilon_{\rm{tot}}=n^\epsilon_c+n^\epsilon_b$ & -  \\
				\hline
			\end{tabular}
		\end{center}
	\end{table}	
	\begin{table}[tbh!]
		\tabcolsep7.5pt
		\caption{Currents}
		\label{tab3}
		\begin{center}
			\begin{tabular}{@{}l|c|c|c@{}}
				\hline
				&Particle Current &Momentum Flux & Energy Current\\
				\hline
				Electron &$\vec{j}_+=\int_{\vec{k}} \partial_{\vec{k}}\epsilon_+(\vec{x},\vec{k},  t)\; f_+$& $ \vec{\vec{\Pi}}^+=\int_{\vec{k}}\vec{k} \partial_{\vec{k}}\epsilon_+(\vec{x},\vec{k},  t)f_{+} $ & $\vec{j}^\epsilon_+=\int_{\vec{k}} \partial_{\vec{k}}\epsilon_+(\vec{x},\vec{k},  t)\epsilon_+(\vec{x},\vec{k},  t)\; f_+$ \\
				Hole  &$\vec{j}_-=\int_{\vec{k}} \partial_{\vec{k}}\epsilon_-(\vec{x},\vec{k},  t)\; f_-$&$\vec{\vec{\Pi}}^-=\int_{\vec{k}}\vec{k} \partial_{\vec{k}}\epsilon_-(\vec{x},\vec{k},  t) f_{-}$ & $\vec{j}^\epsilon_-=\int_{\vec{k}} \partial_{\vec{k}}\epsilon_-(\vec{x},\vec{k},  t)\epsilon_-(\vec{x},\vec{k},  t)\; f_- $ \\
				Charge &$\vec{j}_c=-e\left(\vec{j}_++\vec{j}_-\right)$ &$\vec{\vec{\Pi}}^{c}=\vec{\vec{\Pi}}^{+}+\vec{\vec{\Pi}}^{-} $ & $\vec{j}^\epsilon_c=\vec{j}^\epsilon_++\vec{j}^\epsilon_-$ \\
				Imbalance &$\vec{j}_{\rm{imb}}=-e\left(\vec{j}_+-\vec{j}_-\right) $& $\vec{\vec{\Pi}}^{\rm{imb}}=\vec{\vec{\Pi}}^{+}-\vec{\vec{\Pi}}^{-} $&$\vec{j}^\epsilon_c=\vec{j}^\epsilon_+-\vec{j}^\epsilon_-$ \\
				Boson &$\vec{j}_b= \int_{\vec{k}} \partial_{\vec{k}}\omega(\vec{x},\vec{k},  t)\; b$& $\vec{\vec{\Pi}}^b=\int_{\vec{k}}\vec{k} \partial_{\vec{k}}\omega(\vec{x},\vec{k},  t) \;b  $ &  $\vec{j}^\epsilon_b=\int_{\vec{k}} \partial_{\vec{k}}\omega(\vec{x},\vec{k},  t)\omega(\vec{x},\vec{k},  t)\; b$ \\ Total &-& $\vec{\vec{\Pi}}^{\rm{tot}}=\vec{\vec{\Pi}}^{c}+\vec{\vec{\Pi}}^{b} $ & $\vec{j}^\epsilon_{\rm{tot}}=\vec{j}^\epsilon_c+\vec{j}^\epsilon_b$ \\
				\hline
			\end{tabular}
		\end{center}
	\end{table}
\end{center}
\end{widetext}

\subsubsection{Thermo-electric transport and hydrodynamics}

The theory of hydrodynamics describes the slow relaxation of conserved quantities to thermal equilibrium. So, the underlying assumption is that there are some conserved quantities: charge, momentum, and energy (and possibly more). Furthermore, there is a fast relaxation process that equilibrates the system without relaxing one of those conserved quantities. This has an immediate consequence: to leading order, there are no momentum-relaxing processes. Consequently, when momentum is excited in the system, it cannot decay. This is problematic in the description of a standard bulk transport experiment: if an electric field accelerates charges but their momentum cannot decay, the conductivity becomes infinite.
So, to describe transport phenomena, relaxation processes have to be included.
In the framework of hydrodynamics this happens by including first-order corrections to the constitutive relations, which introduces conductivities as free parameters~\cite{Kovtun2012,Mueller2008,Lucas2018}. 
There is an alternative: starting from the Boltzmann equations and specifying the collision integrals, one has to solve the resulting integro-differential equations, while including the applied field as a perturbation~\cite{Fritz2008a,Fritz2008b,Arnold2000,Lifshitz1981}.
This alternative approach gives direct access to conductivities and does not leave them unspecified. In this paper, we take the approach based on Boltzmann equations.
However, instead of solving the Boltzmann equation explicitly, which usually requires sophisticated numerics, we introduce relaxation times in the scattering integral. 
This corresponds to a one-mode approximation of the scattering integral. 
It is well known that the relaxation-time approximation can violate conservation laws. 
So, we carefully construct the approximation to ensure that all conservation laws discussed in the preceding section are always obeyed. This still leaves a number of unspecified parameters. We find that there is an additional constraint that we can use to restrict the number of free parameters further: it is the Onsager reciprocal relation~\cite{Onsager1931}. The set of conservation laws and Onsager relation allows to mimic a much more sophisticated numerical solution of the Boltzmann equation with surprising precision. This has one major advantage over a full numerical solution (besides being much less demanding): it has very few free parameters that can be easily compared to experiments.

\subsubsection{Relaxation-time approximation}\label{sec:relaxationtime}

In the subsequent section, we \lq solve\rq\  the Boltzmann equations in order to determine the thermo-electric transport coefficients of the hybrid system of electrons, holes, and bosons in the presence of an external electric field $\vec{E}$ and a temperature gradient $\partial_{\vec{x}}T$ within linear-response theory. The Boltzmann equation is too complicated to be solved exactly. The reason is that the collisional terms are integrals involving the distribution functions themselves. Usually, there is no analytical solution, apart from the equilibrium ones. In the subsequent discussion, we consider near-equilibrium transport phenomena. This allows to linearize the distribution functions according to $f_{\pm} \approx f^0_{\pm}+\delta f_{\pm}$ and $b \approx b^0+\delta b$ around their equilibrium forms, where the deviation of the distribution function $\delta f_\pm$ and $\delta b$ are of linear order in the external fields. The solution of the linearized coupled Boltzmann equations is a standard but still technical exercise that eventually has to resort to mode-expansion and numerics~\cite{Ziman2000}. Going through those technical steps is beyond the scope of this article and also obscures the physics a bit. Instead, we perform the relaxation-time approximation~\cite{Lifshitz1981}. In this approximation, all the collisional processes are summarized in one quantity, the relaxation time $\tau$. This approach usually has a problem that it does not necessarily respect conservation laws. The way we set up the relaxation-time approximation, however, is such that it does respect the conservation laws. Furthermore, we checked that it reproduces the qualitative features of the actual numerical solutions. More systematic approaches have been presented across several places in the literature.
The linearized Boltzmann equations in the relaxation-time approximation read
\begin{widetext}
\begin{eqnarray}\label{eq:coupledBoltzmannrelaxtime}
\partial_t \delta f_+ - \frac{\epsilon_+-\mu}{T}  \partial_{\vec{x}}T \cdot \partial_{\vec{k}}f^0_+  - e \vec{E} \cdot \partial_{\vec{k}}f^0_+&=& -\frac{\delta f_+}{\tau_{+}}+\frac{\delta f_-}{\tau_{-}} -\frac{\delta f_+}{\tau^{\rm{dis}}_+}-\frac{\delta f_+}{\tau_{+b}}+\frac{\delta b}{\tau_{b+}} \;, \nonumber \\
\partial_t \delta f_-  - \frac{\epsilon_--\mu}{T}\partial_{\vec{x}}T \cdot \partial_{\vec{k}}f^0_- - e \vec{E}\cdot \partial_{\vec{k}}f^0_-&=& -\frac{\delta f_-}{\tau_{-}}+\frac{\delta f_+}{\tau_{+}} -\frac{\delta f_-}{\tau^{\rm{dis}}_-}-\frac{\delta f_-}{\tau_{-b}}+\frac{\delta b}{\tau_{b-}} \;, \nonumber \\ \partial_t \delta b - \frac{\omega}{T}\partial_{\vec{x}}T \cdot \partial_{\vec{k}}b^0 &=&-\frac{\delta b}{\tau_{b+}} +\frac{\delta f_+ }{\tau_{+b}}-\frac{\delta b}{\tau_{b-}} +\frac{\delta f_- }{\tau_{-b}}\;.
\end{eqnarray}
\end{widetext}
We have introduced a set of relaxation times which all play different physical roles. The relaxation times $1/\tau_{+}$ and $1/\tau_{-}$ refer to electron-hole drag, mediated by interactions, $1/\tau^{\rm{dis}}_{\pm}$ refers to disorder scattering for electrons and holes, respectively, whereas $1/\tau_{b+}$, $1/\tau_{b-}$, $1/\tau_{+b}$, and $1/\tau_{-b}$ refer to drag between electrons, holes, and bosons, and vice versa. For simplicity, we henceforth assume $1/\tau^{\rm{dis}}_+=1/\tau^{\rm{dis}}_-=1/\tau^{\rm{dis}}$, which is justified for near charge-neutral systems.

\section{Part A: Electron-hole plasmas}\label{sec:electronholeplasma}

We have set up the coupled system of Boltzmann equations of electron, holes, and bosons within the relaxation-time approximation.
Let us adopt this approach in analyzing the thermo-electric response of a weakly interacting charged Dirac electron-hole plasma. The prime representative of the class of electron-hole plasmas is graphene close to its charge neutrality point. However, all Dirac-type systems, bilayer graphene~\cite{Nam2017}, and even semiconductors at elevated temperatures fall into this category~\cite{Jaoui2021}; albeit with minor modifications. 

\begin{figure}
\includegraphics[width=0.48\textwidth]{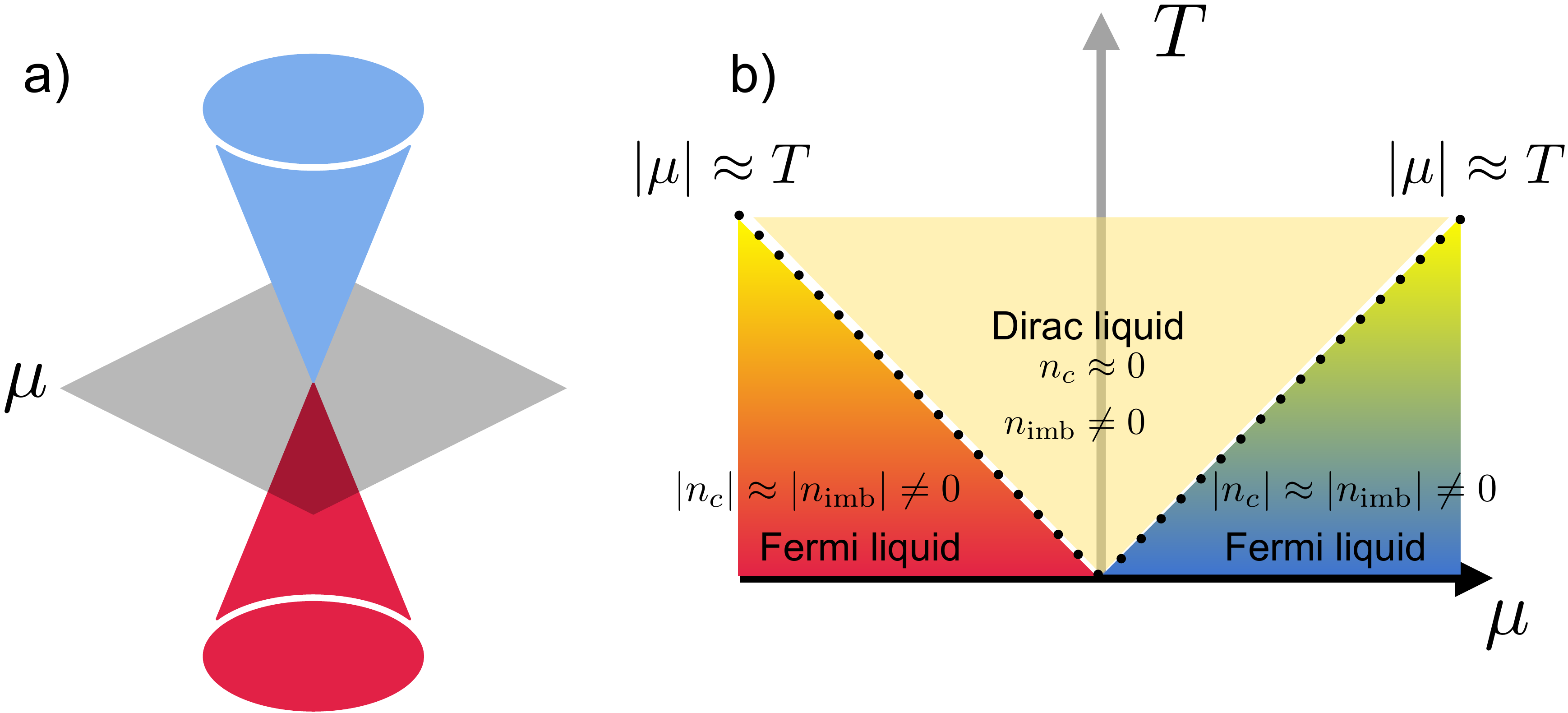}
\caption{a) Schematic of the dispersion relation of graphene near its Dirac point. b) \lq Phase diagram\rq\ of clean graphene at finite temperature. The region above $\mu=0$ is referred to as the Dirac liquid. At $|\mu| \approx T$ it crosses over to a Fermi liquid.}\label{fig4}
\end{figure}

Graphene is a two-dimensional system of carbon atoms on a honeycomb lattice. In its undoped state, it is neither a metal nor an insulator, but a semimetal~\cite{CastroNeto2009,katsnelson2020}. As such, it has a vanishing density of states at zero energy, but it is linear in energy everywhere else. This originates from the low-energy band structure, shown in Fig.~\ref{fig4} a). Two bands touch in isolated points in the Brillouin zone. In the vicinity of these points, the system can effectively be described by the massless Dirac equation. Consequently, the spectrum is linear in momentum according to $\epsilon_{\pm}=\pm v_F |\vec{k}|$, where $+$ refers to electrons, $-$ to holes, and $v_F$ is the Fermi velocity. More details about graphene can be found in the literature \cite{CastroNeto2009,katsnelson2020} but are not necessary here.

The key insight concerning the plasma character of charge-neutral graphene came in a paper by Sheehy and Schmalian in 2007~
\cite{Sheehy2007}. The essence is summarized in Fig.~\ref{fig4} b). It shows the `phase diagram' of graphene as a function of the chemical potential $\mu$ ($x$-axis) and temperature $T$ ($y$-axis). The chemical potential controls the filling of the Dirac cones: the charge density $n_c\propto \mu^2$. Consequently, at $\mu=0$ we have $n_c=0$. However, there are still excitations at nonzero temperature $T$. This is captured by the imbalance density $n_{\rm{imb}}\propto T^2$ introdueced in Table \ref{tab2}. The physics behind this density is that there is a thermal cloud of electrons and one of holes, both of equal density, which ensures $n_c=0$. The nonzero-temperature region above $\mu$ has been dubbed the `Dirac liquid' and it has thermodynamic properties that are very different from Fermi liquids. The crossover region is defined by the condition $|\mu| \approx T$ (up to renormalizations due to interactions). For $|\mu| \gg T$, the system behaves like a Fermi liquid of electron or hole type. This discussion is not only valid in graphene, but in any Dirac-type two-band system; including bilayer graphene. If the temperature is larger than the respective gap, it even applies to semiconductors.

\subsection{Thermo-electric response and Onsager relation}\label{subsec:thermoelectricI}

One of the hallmarks of hydrodynamic behavior in a Dirac plasma is the bulk thermo-electric response. We find two key features experimentally observed in graphene: an interaction dominated electrical conductivity and a strong violation of the Wiedemann-Franz law~\cite{Mahan2000}. 

The thermo-electric response of a system is the combined response of the system to an applied electric field $\vec{E}$ and a temperature gradient $\partial_{\vec{x}}{T}$ across the system. For us, it assumes the form~\cite{Mahan2000}
\begin{eqnarray}\label{eq:tensor}
\left( \begin{array}{c} \vec{j}_c \\ \vec{j}^Q \end{array} \right)=\left( \begin{array}{cc} \sigma & \alpha \\ T \bar{\alpha} & \overline{\kappa} \end{array} \right)\left( \begin{array}{c} \vec{E} \\ -\partial_{\vec{x}}{T}\end{array} \right).
\end{eqnarray} 
Here $\sigma$, $\alpha$, $\bar{\kappa}$ are longitudinal conductivities, as the  currents are in the direction parallel to the perturbations. The thermo-electric coefficient $\alpha$ is used to determine the Seebeck coefficient $S$, $\sigma$ is the electrical conductivity, and $\bar{\kappa}$  is the thermal conductivity. Typically, the thermal conductivity $\kappa$ is defined such that $\vec{j}^Q= -\kappa \partial_{\vec{x}}T$ under the condition of no electric current flow, {\it i.e.}, $\vec{j}_c=0$. Using straighforward algebra, one can show that $\kappa = \bar{\kappa} - T \alpha \sigma^{-1} \overline{\alpha}$.  Moreover, provided that the system respects time-reversal symmetry, the Onsager reciprocal relation~\cite{Onsager1931} requires that 
\begin{equation}
\alpha=\bar{\alpha}.
\label{eq:Onsagerrelation}
\end{equation} Later, we use this relation to establish the relationship between the relaxation times associated with the drag effect in Eq.~\eqref{eq:twofluidboltzmann}. The Boltzmann transport equations for electrons and holes are given by
\begin{widetext}
\begin{eqnarray}\label{eq:twofluidboltzmann}
\partial_t \delta f_+ - \frac{\epsilon_+-\mu}{T} \partial_{\vec{x}}T\cdot  \partial_{\vec{k}}f^0_+  - e \vec{E}\cdot \partial_{\vec{k}}f^0_+&=&-\frac{\delta f_+}{\tau_{+}}+\frac{\delta f_-}{\tau_{-}}-\frac{\delta f_+}{\tau^{\rm{dis}}} \nonumber \\ \partial_t \delta f_-  - \frac{\epsilon_--\mu}{T} \partial_{\vec{x}}T \cdot \partial_{\vec{k}}f^0_- - e \vec{E}\cdot \partial_{\vec{k}}f^0_-&=&-\frac{\delta f_-}{\tau_{-}}+\frac{\delta f_+}{\tau_{+}}-\frac{\delta f_-}{\tau^{\rm{dis}}}\;.
\end{eqnarray}
\end{widetext}
We may determine the transport coefficients from the equations of motion for charge and heat current densities. The equation for electron and hole currents are obtained by integrating each Boltzmann equation over the corresponding group velocity $\vec{v}_\pm = \partial_{\vec{k}}\epsilon_{\pm}(\vec{k})$. In a straightforward way this leads to 
\begin{equation}
	\partial_t \vec{j}_\pm -\frac{\partial_{\vec{x}}T}{T} \cdot \vec{\vec{\mathcal{T}}}_\pm-e\vec{E}\cdot\vec{\vec{\mathcal{E}}}_\pm=-\frac{\vec{j}_\pm}{\tau_\pm}-\frac{\vec{j}_\mp}{\tau_\mp}-\frac{\vec{j}_\pm}{\tau^{\rm{dis}}}\;,
	\label{eq:currentequation}
\end{equation}
where
\begin{eqnarray}
	\vec{\vec{\mathcal{E}}}_\pm &=& \int_{\vec{k}} \vec{v}_\pm(\vec{k})\; \partial_{\vec{k}}f^0_{\pm}(\vec{k})\;,
	\nonumber\\
	\vec{\vec{\mathcal{T}}}_\pm&=& \int_{\vec{k}} \vec{v}_\pm(\vec{k}) \left(\epsilon_{\pm}-\mu\right) \partial_{\vec{k}}f^0_{\pm}(\vec{k})\;.
	\label{eq:EandT}
\end{eqnarray}

The equation of motion for the thermal currents of the electrons and holes can be obtained by integrating the Boltzmann equations over the corresponding product of the group velocities and the energies, $\vec{v}_\pm \left(\epsilon_\pm-\mu\right)$. This leads to  
\begin{equation}
		\partial_t \vec{j}^Q_\pm -\frac{\partial_{\vec{x}}T}{T} \cdot \vec{\vec{\mathcal{S}}}_\pm-e\vec{E}\cdot\vec{\vec{\mathcal{T}}}_\pm=-\frac{\vec{j}^Q_\pm}{\tau_\pm}+\frac{\vec{j}^Q_\mp+2\mu\vec{j}_\mp}{\tau_\mp}-\frac{\vec{j}^Q_\pm}{\tau^{\rm{dis}}},
		\label{eq:heatcurrentequation}
\end{equation}
where the thermal current is defined as
\begin{equation}
	\vec{j}^Q_\pm = \int_{\vec{k}}\vec{v}_\pm(\vec{k})\left(\epsilon_\pm(\vec{k})-\mu\right)\delta f_\pm\;,
\end{equation} 
whereas we have
\begin{equation}
	\vec{\vec{\mathcal{S}}}_\pm = \int_{\vec{k}} \vec{v}_\pm(\vec{k}) \left(\epsilon_{\pm}-\mu\right)^2 \partial_{\vec{k}}f^0_{\pm}(\vec{k})\;.
	\label{eq:S}
\end{equation}
The total charge and thermal current densities are given by
\begin{equation}
	\vec{j}_c = -e\left(\vec{j}_+ +\vec{j}_-\right)\;,
	\label{eq:totalchargecurrent}
\end{equation}
and
\begin{equation}
	\vec{j}^Q = \vec{j}^Q_+ +\vec{j}^Q_-\;,
		\label{eq:totalheatcurrent}
\end{equation}
respectively. We are interested in the steady-state situation, {\it i.e.},  $\partial_t \vec{j}_\pm=0$ and $\partial_t \vec{j}^Q_\pm=0$. For that, it is convenient to cast Eqs.~\eqref{eq:currentequation} and~\eqref{eq:heatcurrentequation} in a matrix form according to
\begin{widetext}
\begin{equation}
	\begin{pmatrix}
		\frac{1}{\tau_+}+\frac{1}{\tau^{\rm{dis}}}&\frac{1}{\tau_-}&0&0\\\frac{1}{\tau_+}&\frac{1}{\tau_-}+\frac{1}{\tau^{\rm{dis}}}&0&0\\0&-2\mu\frac{1}{\tau_-}&\frac{1}{\tau_+}+\frac{1}{\tau^{\rm{dis}}}&-\frac{1}{\tau_-}\\-2\mu\frac{1}{\tau_+}&0&-\frac{1}{\tau_+}&\frac{1}{\tau_-}+\frac{1}{\tau^{\rm{dis}}}
	\end{pmatrix}
	\begin{pmatrix}
		\vec{j}_+\\\vec{j}_- \\ \vec{j}_+^Q \\ \vec{j}_-^Q
	\end{pmatrix} = \frac{\partial_{\vec{x}}T}{T}\cdot\begin{pmatrix}
	\vec{\vec{\mathcal{T}}}_+\\  \vec{\vec{\mathcal{T}}}_-\\ \vec{\vec{\mathcal{S}}}_+ \\ \vec{\vec{\mathcal{S}}}_-
\end{pmatrix}
+ e\vec{E} \cdot
	\begin{pmatrix}
		 \vec{\vec{\mathcal{E}}}_+\\  \vec{\vec{\mathcal{E}}}_- \\  \vec{\vec{\mathcal{T}}}_+\\  \vec{\vec{\mathcal{T}}}_-
	\end{pmatrix}.
\end{equation}
\end{widetext}
It is straightforward to solve this matrix equation for electrical and heat currents. By substituting the result into Eqs.~\eqref{eq:totalchargecurrent} and \eqref{eq:totalheatcurrent}, we find that the longitudinal coefficients take the form: 
\begin{equation}
	\sigma = -e^2\frac{\left(\mathcal{E}_{+}\left(\frac{1}{\tau^{\rm{dis}}}+\frac{1}{\tau_-}-\frac{1}{\tau_+}\right)+\mathcal{E}_{-}\left(\frac{1}{\tau^{\rm{dis}}}+\frac{1}{\tau_+}-\frac{1}{\tau_-}\right)\right)}{\frac{1}{\tau^{\rm{dis}}}\left(\frac{1}{\tau^{\rm{dis}}}+\frac{1}{\tau_+}+\frac{1}{\tau_-}\right)}\;,
	\label{eq:sigma}
\end{equation}
\begin{equation}
	\bar{\kappa} =-\frac{1}{T} \frac{\left(\mathcal{S}_{+}+\mathcal{S}_{-}\right)\left(\frac{1}{\tau^{\rm{dis}}}+\frac{1}{\tau_-}+\frac{1}{\tau_+}\right)-2\mu\left(\frac{\mathcal{T}_{+}}{\tau_+}+\frac{\mathcal{T}_{-}}{\tau_-}\right)}{\frac{1}{\tau^{\rm{dis}}}\left(\frac{1}{\tau^{\rm{dis}}}+\frac{1}{\tau_+}+\frac{1}{\tau_-}\right)}\;,
	\label{eq:kappa}
\end{equation}
\begin{equation}
	\alpha= 
	\frac{e}{T}\frac{\mathcal{T}_{+}\left(\frac{1}{\tau^{\rm{dis}}}+\frac{1}{\tau_-}-\frac{1}{\tau_+}\right)+\mathcal{T}_{-}\left(\frac{1}{\tau^{\rm{dis}}}+\frac{1}{\tau_+}-\frac{1}{\tau_-}\right)}{\frac{1}{\tau^{\rm{dis}}}\left(\frac{1}{\tau^{\rm{dis}}}+\frac{1}{\tau_+}+\frac{1}{\tau_-}\right)}\;,
	\label{eq:alpha}
\end{equation}
\begin{equation}
	\bar{\alpha} =  	\frac{e}{T}\frac{\left(\mathcal{T}_{+}+\mathcal{T}_{-}\right)\left(\frac{1}{\tau^{\rm{dis}}}+\frac{1}{\tau_+}+\frac{1}{\tau_-}\right)-2\mu\left(\frac{\mathcal{E}_{+}}{\tau_+}+\frac{\mathcal{E}_{-}}{\tau_-}\right)}{\frac{1}{\tau^{\rm{dis}}}\left(\frac{1}{\tau^{\rm{dis}}}+\frac{1}{\tau_+}+\frac{1}{\tau_-}\right)}\;.
\end{equation}
Here,
 $\mathcal{E}_{\pm}$, $\mathcal{T}_{\pm}$, and $\mathcal{S}_{\pm}$ are the main diagonal elements of the second-rank tensors, so $\vec{\vec{\mathcal{E}}}_{\pm} = \mathcal{E}_\pm \vec{\vec{\mathbb{1}}}$, $\vec{\vec{\mathcal{T}}}_{\pm}= \mathcal{T}_\pm \vec{\vec{\mathbb{1}}}$, and $\vec{\vec{\mathcal{S}}}_{\pm}= \mathcal{S}_\pm \vec{\vec{\mathbb{1}}}$. They are evaluated from Eqs.~\eqref{eq:EandT} and \eqref{eq:S}. For Dirac fermions with a linear energy dispersion, we find that 
\begin{eqnarray}
	\mathcal{E}_{\pm} &=& \frac{NT}{4\pi}\text{Li}_{1}\left(- e^{\pm \mu/T}\right)\;\nonumber\\
	\mathcal{T}_{\pm}&=& \pm \frac{NT^2}{2\pi}\text{Li}_2\left(-e^{\pm\mu/T}\right) - \frac{NT\mu}{4\pi} \text{Li}_{1}\left(-e^{\pm\mu/T}\right)\;,
	\nonumber\\
	\mathcal{S}_{\pm}&=& -\frac{NT}{4\pi}\Big[-3T^2\text{Li}_3\left(-e^{\pm\mu/T}\right)\pm4T\mu\text{Li}_{2}\left(-e^{\pm \mu/T}\right)\nonumber\\&&-\mu^2\text{Li}_1\left(-e^{\pm \mu/T}\right)\Big]\ ,
	\label{eq:ExxTxxSxx}
\end{eqnarray}
where $\text{Li}_n$ is the polylogarithmic function of order $n$ and $N$ denotes the number of Dirac fermion flavors ($N=4$ in the case of graphene, accounting from spin and valley degeneracy).
It can be verified that, for the Onsager reciprocal relation in Eq. (\ref{eq:Onsagerrelation}) to hold, the relaxation time for the electron-hole drag must satisfy the relation  
\begin{equation}
	\frac{1}{\tau_-}\left( \mathcal{T}_{-}+\mu\mathcal{E}_{-}\right) = - \frac{1}{\tau_+}\left( \mathcal{T}_{+}+\mu\mathcal{E}_{+}\right).
	\label{eq:relationbetweentauplusandtuaminus}
\end{equation}
Therefore, we can choose
\begin{eqnarray}
	\frac{1}{\tau_+} &=& \frac{1}{\tau_0}\frac{ \mathcal{T}_{-}+\mu\mathcal{E}_{-}}{\left( \mathcal{T}_{-}+\mu\mathcal{E}_{-}\right)-\left( \mathcal{T}_{+}+\mu\mathcal{E}_{+}\right)}\;,\nonumber\\
	\frac{1}{\tau_-} &=& -\frac{1}{\tau_0}\frac{\mathcal{T}_{+}+\mu\mathcal{E}_{+}}{\left( \mathcal{T}_{-}+\mu\mathcal{E}_{-}\right)-\left( \mathcal{T}_{+}+\mu\mathcal{E}_{+}\right)}\;,
	\label{eq:taudrag}
\end{eqnarray}
where $\tau_0$ is a relaxation time that has to be determined from a Boltzmann equation. All the quantities in Eqs.~\eqref{eq:relationbetweentauplusandtuaminus} and~\eqref{eq:taudrag} depend strongly on $\mu/T$; including $\tau_0$. Since the system at charge neutrality is a quantum critical system, the relaxation time there is of order $1/T$, saturating at the Planckian limit. This fixes $1/\tau_0 \propto T$. Upon moving away from charge neutrality, all drag effects are strongly suppressed. We demonstrate this explicitly in Appendix~\ref{sec:scatteringtime}. For practical calculations using the relaxation time approximation, we used the interpolation function $1/\tau_0=0.6 T \exp(-\abs{\mu}/T)$. This function shows good qualitative agreement with the actual functions calculated in Appendix~\ref{sec:scatteringtime}. An important consequence of this is that at large $|\mu|/T$, everything is disorder dominated.
\begin{figure}
	\centering
	\includegraphics[scale=0.7]{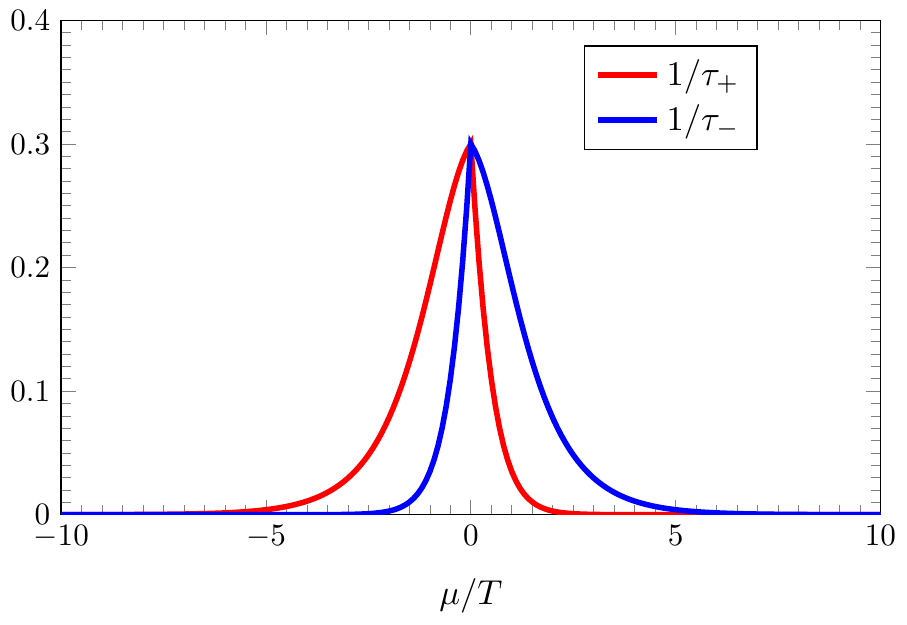}
	\caption{The relaxation time for the electron-hole drag near the charge neutrality point $\mu=0$ calculated from Eq. (\ref{eq:taudrag}). The chemical potential and the relaxation time are scaled in units of $T$. We use $1/\tau_0=0.6T\exp(-\abs{\mu}/T)$ here for illustrative purposes.}
	\label{fig:taudrag}
\end{figure}
In Fig. \ref{fig:taudrag}, we show the relaxation time for the electron-hole drag.
Since clean graphene near the Dirac point is a quantum critical system, this implies that the relaxation time for the drag effect is of order $1/T$; saturating at the Planckian limit.

\begin{figure}
		\centering
		\includegraphics[scale=0.7]{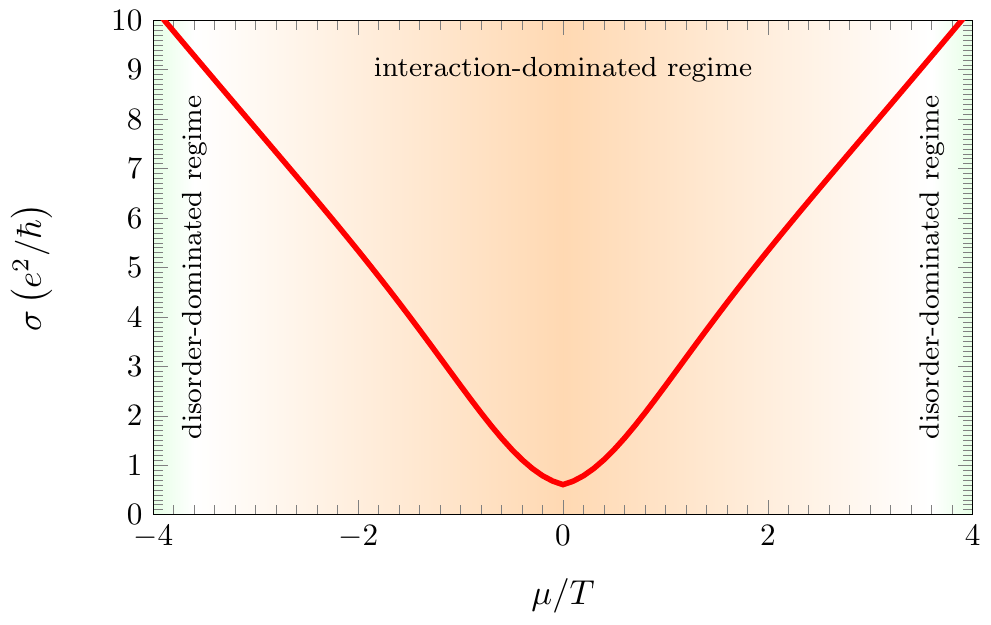}
		\includegraphics[scale=0.7]{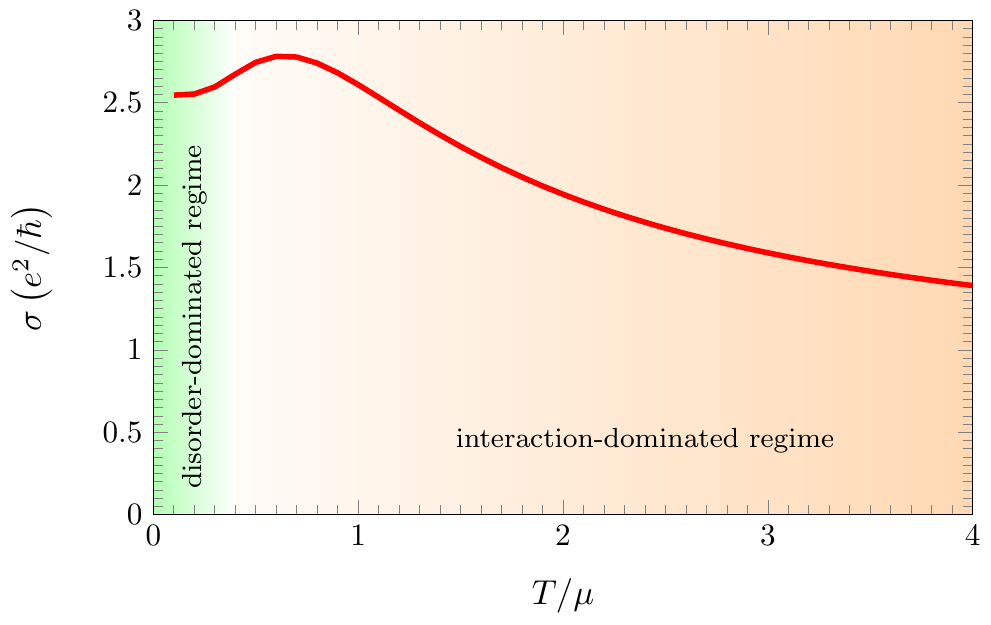}
		\caption{ The upper plot shows the electrical conductivity calculated from Eq. (\ref{eq:sigma}) versus chemical potential. The value of $\mu$ is plotted in units of $T$. In this plot, we use $1/\tau_0=0.6T\exp(-\abs{\mu}/T)$ and $\tau^{\rm{dis}}=8/T$ The lower plot shows the electrical conductivity versus temperature in units of $\mu$. In this plot, we use $1/\tau_0=0.6T\exp(-\abs{\mu}/T)$ and $\tau^{\rm{dis}}=8/\mu$.}
		\label{fig:sigma}
\end{figure}

\subsubsection{The electrical conductivity}
\begin{figure}[h]
	\includegraphics[width=0.48\textwidth]{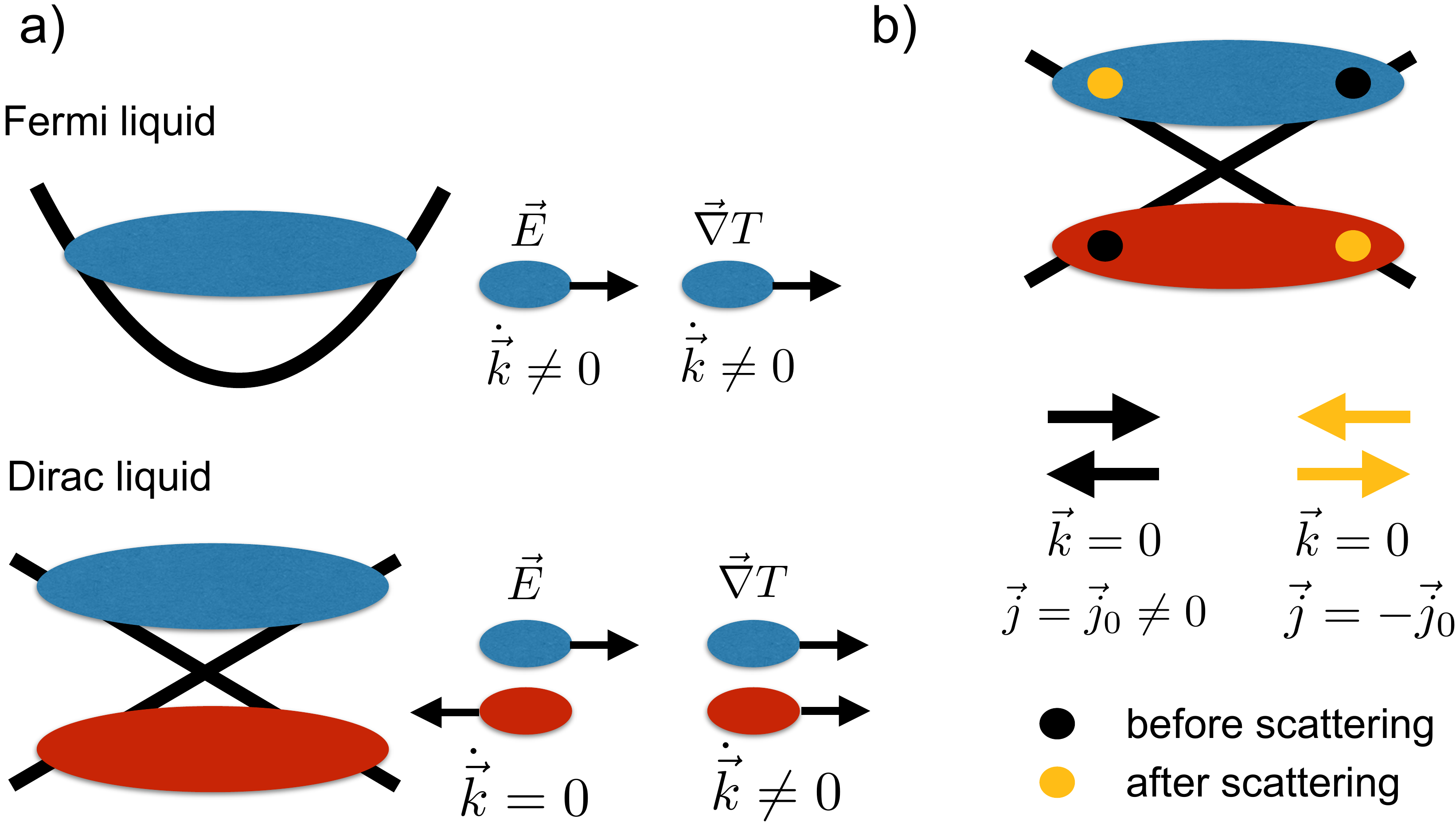}
	\caption{a) In a Fermi liquid, an applied electric field as well as a temperature gradient excite nonzero momentum. In the Dirac liquid, a temperature gradient excites a nonzero momentum, whereas an electric field does not. b) In the Dirac liquid, momentum and current decouple. One can relax current without relaxing momentum.}
	\label{fig5}
\end{figure}

In the upper panel of Fig. \ref{fig:sigma}, we plot the electrical conductivity as a function of the chemical potential near the neutrality point.
A key signature of the electron-hole plasmas sits in their electrical conductivity and becomes most apparent at charge neutrality $\mu=0$. It revolves around a somewhat paradoxically looking situation. The system has a total charge zero, $n_c=0$. Nevertheless, the d.c. conductivity in the clean limit is finite. From Eq.~\eqref{eq:sigma}, we find that the electrical conductivity at charge neutrality is given by 
\begin{eqnarray}\label{eq:conductivitycrit}
	\sigma(\mu=0)= -e^2\frac{\mathcal{E}_{+}+\mathcal{E}_{-}}{\frac{1}{\tau^{\rm{dis}}}+\frac{2}{\tau_+}}\ ,
\end{eqnarray}
where, as depicted in Fig. \ref{fig:taudrag}, the inverse drag scattering times for electron and hole are identical at charge neutrality, {\it i.e.}, $1/\tau_-=1/\tau_+$, and, according to Eq.~(\ref{eq:ExxTxxSxx}),  $\mathcal{E}_{+}+\mathcal{E}_{-} \neq 0$. This implies that the electrical conductivity is finite, even in the absence of disorder, {\it i.e.}, for $1/\tau^{\rm{dis}}=0$. 
The key to understanding it is that while the charge density $n_c=0$, the imbalance density $n_{\rm{imb}}\neq 0$. This is in stark contrast to a conventional Fermi liquid in which there is essentially no distinction between the two densities. Importantly, at nonzero temperature, there are two thermal clouds of equal density, one of electrons and one of holes. An applied electric field will pull the different types of charge carriers into opposite directions. One consequence of this is that while electrons and holes are pulled apart, the total momentum of the system remains zero. This is different in a Fermi liquid, in which an applied field automatically excites momentum. Since there is no net momentum induced, it is also no issue that there are no impurities to relax the momentum. However, there still is a mechanism which `glues' the electrons and holes; namely, drag. This is sufficient to establish a finite electric current, even in the absence of disorder. A graphical illustration of this situation in shown in Fig.~\ref{fig5}.

With a bit of algebra, one can show that Eq.~(\ref{eq:sigma}) takes the form
\begin{equation}
	\sigma=\sigma(\mu=0)+\frac{e^2\left( \mathcal{E}_{-}-\mathcal{E}_{+}\right)\left(\frac{1}{\tau_-}-\frac{1}{\tau_+}\right)}{\frac{1}{\tau^{\rm{dis}}}\left(\frac{1}{\tau^{\rm{dis}}}+\frac{1}{\tau_+}+\frac{1}{\tau_-}\right)}\;.
\end{equation}
The second part becomes singular in the limit of zero disorder: upon tuning away from charge neutrality, $\mu\neq0$, both $\mathcal{E}_{-}-\mathcal{E}_{+}$ and $1/\tau_--1/\tau_+$ are non-zero. This implies that the electrical conductivity is not finite in the absence of disorder ($1/\tau^{\rm{dis}} \to 0$), but diverges, as one would expect. This situation is  qualitatively discussed in Fig.~\ref{fig5}. 

\subsubsection{The thermo-electric coefficient}

\begin{figure}[t]
	\centering	\includegraphics[scale=0.7]{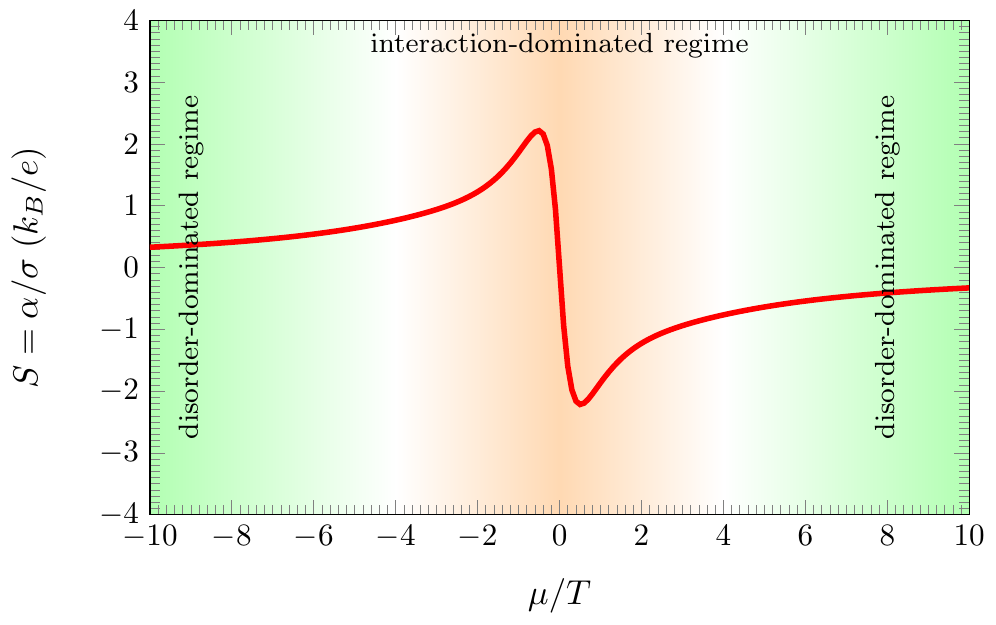}
	\includegraphics[scale=0.7]{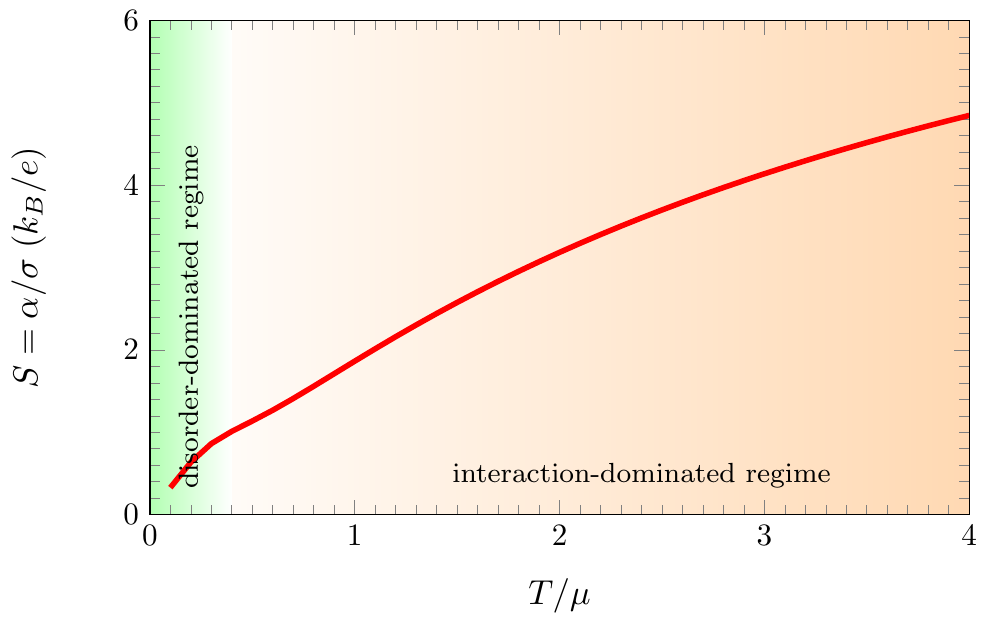}
	\caption{ The upper plot shows the Seebeck coefficient as a function of chemical potential in units of $T$ and the lower plot shows the Seebeck coefficient as a function of temperature in units of $\mu$. The values of $\tau_0$ and $\tau^{\rm{dis}}$ are the same as in Fig. \ref{fig:sigma}.}
	\label{fig:Seebeck}
\end{figure}
In the upper panel of Fig. \ref{fig:Seebeck}, we plot the Seebeck coeeficient, which is the ratio of the thermo-electric cofficient $\alpha$ to the electrical conductivity $\sigma$, as a function of the chemical potential near the neutrality point. A main signature of the electron-hole plasmas is that at the charge neutrality, the Seebeck coefficient vanishes because
\begin{equation}
	\alpha(\mu=0)=\frac{e}{T}\frac{\mathcal{T}_{+}+\mathcal{T}_{-}}{\frac{1}{\tau^{\rm{dis}}}+\frac{2}{\tau_+}} = 0.
\end{equation}
The key to understanding it is that when the Dirac plasma is exposed to a temperature gradient, both thermal clouds of electrons and holes of equal concentration move in the same direction. As a result, $\vec{j}_c = 0$, which implies $\alpha = 0$. 
\subsubsection{The heat conductivity}
\begin{figure}[t]
	\centering	\includegraphics[scale=0.7]{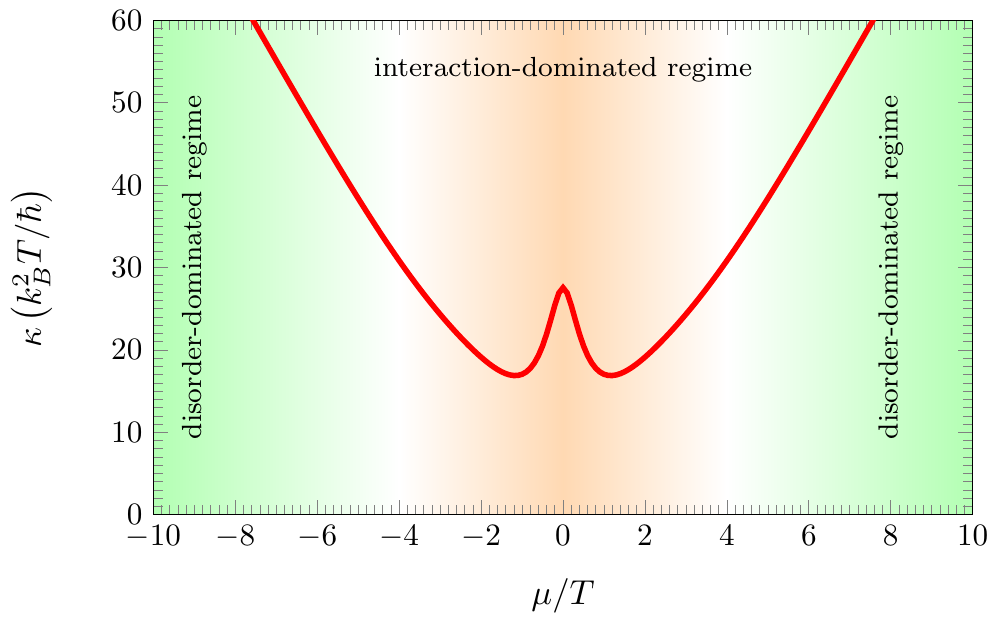}
	\includegraphics[scale=0.7]{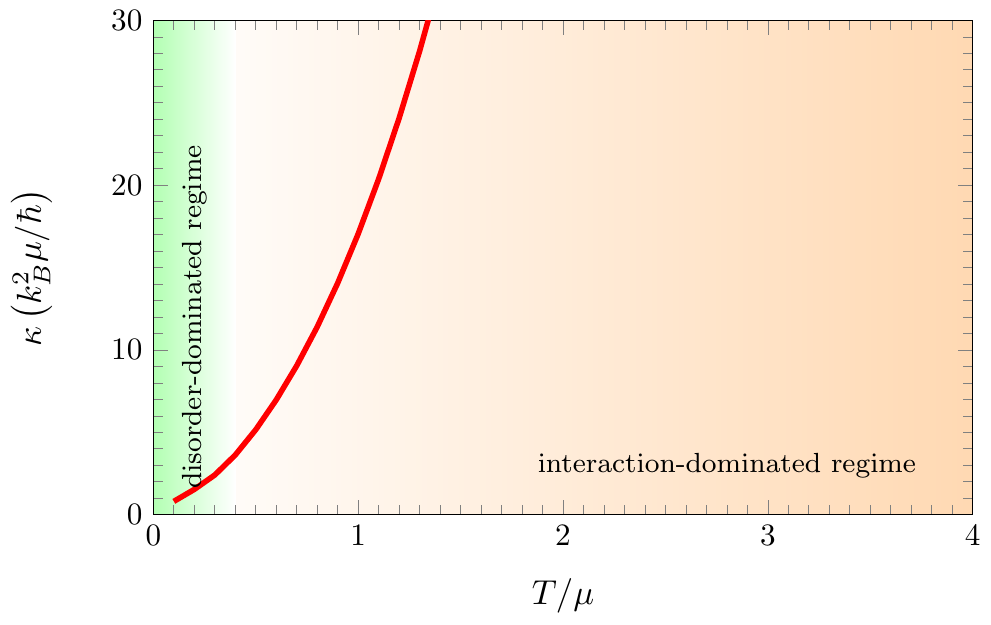}
	\caption{ The upper plot shows the thermal conductivity calculated from $\kappa= \bar{\kappa}- T\alpha^2/\sigma$ versus the chemical potential. The lower plot shows the thermal conductivity versus temperature in units of $\mu$. The values of $\tau_0$ and $\tau^{\rm{dis}}$ are the same as in Fig. \ref{fig:sigma} .}
	\label{fig:thermalconductivity}
	\label{}
\end{figure}
\begin{figure}[t]
	\centering	\includegraphics[scale=0.7]{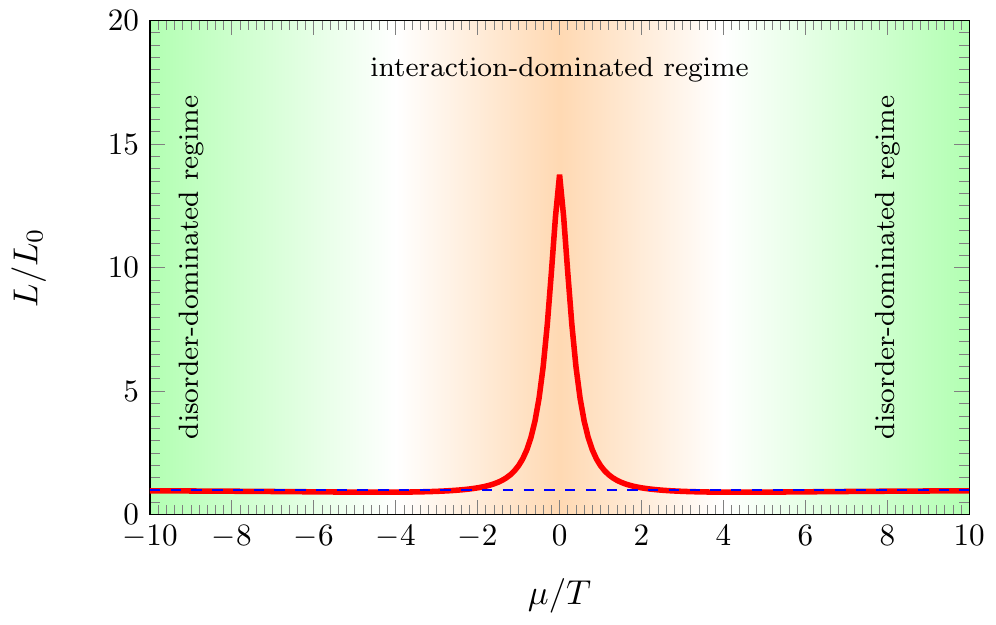}
	\includegraphics[scale=0.7]{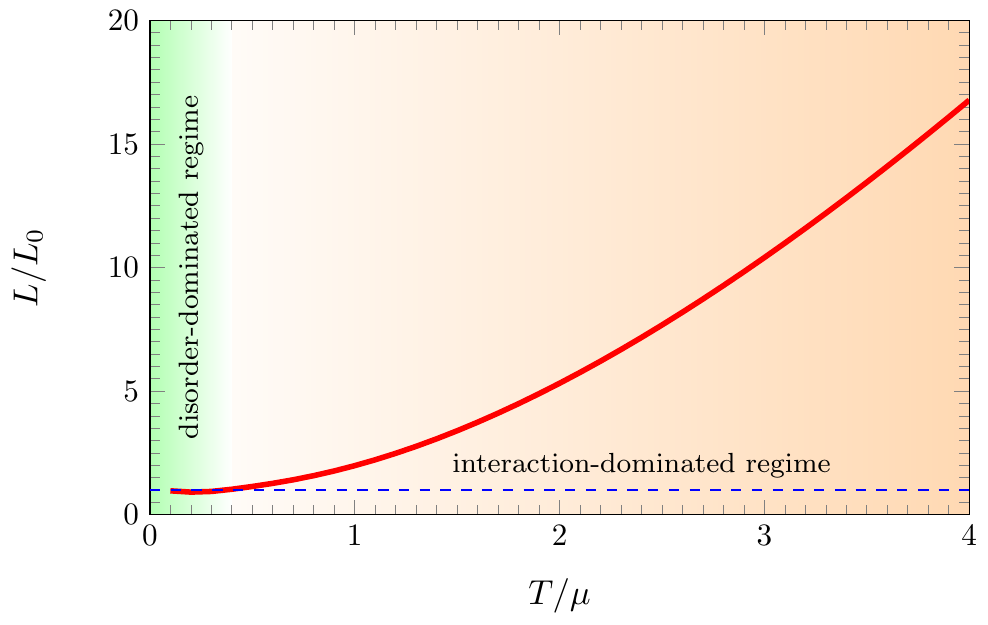}
	\caption{ The upper plot shows the Lorenz ratio versus the chemical potential $\mu$ in units of $T$. 
The lower plot shows the Lorenz ratio versus temperature in units of $\mu$. The values of $\tau_0$ and $\tau^{\rm{dis}}$ are the same as in Fig. \ref{fig:sigma}.}
	\label{fig:Wiedemann-FranzratioA}
\end{figure}
Here, we consider the response $\vec{j}^Q=-\kappa \partial_{\vec{x}}T$, which is the heat conductivity in the absence of current flow, where the longitudinal thermal conductivity  $\kappa = \bar{\kappa} - T\alpha^2/\sigma$. In Fig. \ref{fig:thermalconductivity}, we plot the thermal conductivity of two-dimensional Dirac electrons close to the neutrality point. At the charge neutrality point, we find that
\begin{equation}
	\kappa(\mu=0) = -\frac{1}{T}\left( \mathcal{S}_{+}+\mathcal{S}_{-}\right) \tau^{\rm{dis}} \;.
\end{equation}
Contrary to the electrical conductivity, the thermal conductivity decays from disorder and cannot relax via drag at the charge neutrality point. The reason is simple: the two thermal clouds, electrons and holes, are both dragged into the same direction. Consequently, a net momentum is induced. This implies that momentum has to be relaxed to establish a stationary state with a finite conductivity. This can only be achieved by scattering from impurities; meaning that the heat conductivity diverges in the clean system. This is again explained in Fig.~\ref{fig5}.
\subsubsection{The Wiedemann-Franz ratio}
An important quantity in the study of metals is the Wiedemann-Franz ratio. This was already established in 1853~\cite{WiedemannFranz1853}, based upon the observation that for a variety of metals, the ratio $\kappa/(T\sigma)$ tends to a constant, universal value at low temperatures, called the Lorenz number~\cite{Mahan2000}. It was experimentally found that the Lorenz number is given by
\begin{eqnarray}
	L=\frac{\kappa}{T\sigma}=L_0= \frac{\pi^2}{3} \left( \frac{k_B}{e}\right)^2\;.
\end{eqnarray}
Whether a system tends to this value or not is still often taken as an empirical evidence of whether the system is a Fermi liquid or not. The standard understanding is that both heat and electrical currents are transported by the same type of (quasi-)particle. Additionally, both heat and electrical current undergo the same relaxational mechanism. In the case of a standard metal this means that both heat and electrical current are limited by the same scattering time $\tau^{\rm{dis}}$.
In Fig.~\ref{fig:Wiedemann-FranzratioA} , we plot the Lorenz number for the two-dimensional Dirac fluid around the charge neutrality point. At the charge neutrality point, we find 
\begin{eqnarray}\label{eq:WF}
	L/L_0=\frac{\mathcal{S}_{+}+\mathcal{S}_{-}}{ \mathcal{E}_{+}+\mathcal{E}_{-}}  \left( 1+\tau_{\rm{dis}} \frac{\tau_+ + \tau_-}{\tau_+ \tau_-}  \right)   \;.
\end{eqnarray}
Not only does this ratio diverge for a clean system, it is also, in general, not a universal quantity: one should expect a possibly strong violation of the Wiedemann-Franz law close to charge neutrality as well as a strong variation across different samples. However, it is an excellent measure to determine the relative strength of elastic and inelastic scattering in the system. It is important to note that one cannot extend Eq.~\eqref{eq:WF} towards the Fermi liquid regime; it is only valid at charge neutrality.
\subsection{Internal electric field and plasmons}\label{subsec:plasmons}
It is well known that interacting electronic systems host a variety of collective excitations. A particularly important excitation of that type is the plasmon, or charge oscillation~\cite{Pines&Bohm1952,Bohm&Pines1953}. A standard derivation of the plasmon spectrum is based on many-body techniques that employ the so-called random-phase approximation (RPA)~\cite{Mahan2000}. Alternatively, one can derive plasmons from the Boltzmann equations or continuity equations if they are supplemented by the Poisson equation. In the case of two-dimensional systems, most notably graphene, there is a discrepancy between the hydrodynamic version of plasmons and the plasmons from RPA. This difference becomes particularly visible in the vicinity of the charge neutrality point when temperature dominates the chemical potential and $T \gg \mu$. In the extreme limit of zero charge density, $\mu=0$, the hydrodynamic theory predicts the absence of plasmons~\cite{Lucas2018}, whereas RPA predicts the existence of thermal plasmons~\cite{DasSarma2007} with a dispersion that follows
\begin{eqnarray}
	\omega (\vec{k}) \propto \sqrt{T k} \;.
\end{eqnarray}
In this section we show that the inclusion of drag effects via $C_{\pm}^{eh/he}$ allows to describe both limits, the RPA and the hydrodynamic limit, as well as the crossover between the two limits.  The process which mediates between the two limits is the electron-hole momentum drag, which is also responsible for the finite conductivity at charge neutrality, even in the clean limit, see Eq.~\eqref{eq:conductivitycrit}. This observation constitutes the main result of this section. 

\subsubsection{Coulomb interaction}

Coulomb interactions not only play an important role in the equilibration via electron-electron scatterings, it is also responsible for the internal classical Coulomb force between electrons in the system. The internal Coulomb force is self-consistently determined from the Poisson equation. Projected into two dimensions, it reads
\begin{equation}
	 \sqrt{- \partial_{\vec{x}}^2} V = \frac{\delta n_c}{2\epsilon}\ ,
\end{equation}
with $V$ being the Coulomb potential~\cite{Mahan2000}. It has the well known solution
\begin{eqnarray}
	V(\vec{x}) &=& \int d^2\vec{x}' \frac{-e}{4\pi \epsilon} \frac{(n_+(\vec{x}')-n_+^0(\vec{x}'))+(n_-(\vec{x}')-n_-^0(\vec{x}'))}{|\vec{x}-\vec{x}'|} \nonumber \\ &=& \int d^2\vec{x}' \frac{-e}{4\pi \epsilon} \frac{\delta n_+(\vec{x}')+ \delta n_-(\vec{x}')}{|\vec{x}-\vec{x}'|} \;,
\end{eqnarray}
where $n_+^0(\vec{x})$ and $n_-^0(\vec{x})$ are the equilibrium densities of electrons and holes respectively; furthermore, we introduced the deviations from the equilibrium densities $\delta n_\pm(\vec{x}) = n_\pm(\vec{x})-n_\pm^0(\vec{x})$.
The resulting electric field, after Fourier transformation, reads
\begin{equation}\label{eq:field}
	\vec{E}= -\frac{i  \delta n_c  \vec{k}}{2 \epsilon k} \;,
\end{equation}
where the particular momentum dependence is special to two dimensions.

\subsubsection{The linearized hydrodynamic equations}
Our starting point are the linearized equations for charge densities and current densities.
\begin{eqnarray}
	\partial_t \delta n_c + \partial_{\vec{x}} \cdot  \vec{j}_c&=& 0 \;,\nonumber \\  
	\partial_t \vec{j}_c +e^2\vec{E}\cdot\left(\vec{\vec{\mathcal{E}}}_++\vec{\vec{\mathcal{E}}}_-\right)&=&-\frac{\vec{j}_c}{\tau_c}-\frac{\vec{j}_{\text{imb}}}{\tau_{\text{imb}}}\;,\nonumber\\
		\partial_t \vec{j}_{\text{imb}} +e^2\vec{E}\cdot\left(\vec{\vec{\mathcal{E}}}_+-\vec{\vec{\mathcal{E}}}_-\right)&=&0\;,
\end{eqnarray}
where the charge and imbalance densities and currents are defined in Tables \ref{tab2} and \ref{tab3}. Here $1/\tau_c=1/\tau_++1/\tau_-$ and $1/\tau_{\text{imb}} = 1/\tau_+-1/\tau_-$
We can cast this into an eigenvalue problem according to 
\begin{eqnarray}\label{eq:matrix}
	\left( -i\omega \mathbb{1}+\mathcal{D}_{\rm{eff}}\right) \left(  \delta n_c,\;   \vec{j}_c,\;   \vec{j}_{\text{imb}}\right)^T=0\;,
\end{eqnarray}
with 
\begin{eqnarray}
	\mathcal{D}_{\rm{eff}}=\left( \begin{array} {cccc} 0  &i\vec{k}&0  \\
	 -i e^2\left(\mathcal{E}_++\mathcal{E}_-\right)  \vec{k}/(2\epsilon k) & 1/\tau_{c} & 1/\tau_{\text{imb}} \\ -i e^2\left(\mathcal{E}_+-\mathcal{E}_-\right)  \vec{k}/(2\epsilon k)&0  & 0 
	\end{array}\right) \nonumber \\
\end{eqnarray}
and $\omega$ being the eigenvalue (note that $\mathcal{E}_{\pm}<0$). This can be solved in general, however, the solutions are lengthy. In the following, we
consequently discuss two special cases. 
(i) At the charge neutrality point where things simplify significantly. At this point, we have $1/\tau_{\text{imb}} = 0$ and $\mathcal{E}_+-\mathcal{E}_-=0$. The characteristic equation reduces to
\begin{eqnarray}\label{eq:characteristic}
	-i\omega(-i\omega+1/\tau_c) + \frac{e^2}{2\epsilon }|\mathcal{E}_++\mathcal{E}_-|k=0\;.
\end{eqnarray}
The solutions of this equation read
\begin{equation}
	\omega_{\pm} =  -\frac{i}{2\tau_c} \pm \sqrt{\frac{e^2}{2\epsilon }|\mathcal{E}_++\mathcal{E}_-|k-\left(\frac{1}{2\tau_c}\right)^2}.
	\label{eq:plasmonspectrum}
\end{equation}

We observe that the plasmon modes are damped by fermionic drag.  It is important to note that this mode exists even if $n_c=0$ (in analogy with the finite conductivity at the charge neutrality point) since $\mathcal{E}_++\mathcal{E}_- \neq 0 $. For small momenta, {\it i.e.},
\begin{equation}
	\frac{e^2}{2\epsilon }|\mathcal{E}_++\mathcal{E}_-| ) k < \left(\frac{1}{2\tau_c}\right)^2,
\end{equation} 
they turn into relaxational modes with a purely imaginary frequency.

(ii) Further away from the charge neutrality point, both $1/\tau_c$ and $1/\tau_{\text{imb}}$ vanish. In this case, a true hydrodynamic plasmon emerges with the dispersion given by
\begin{equation}
	\omega_\pm=\sqrt{\frac{e^2}{2\epsilon }|\mathcal{E}_++\mathcal{E}_-|k} =\sqrt{  \frac{e^2}{8 \pi \epsilon v_F} \frac{  n_c^2}{n^\epsilon_c+P}k}\;,
	\label{eq:hydroplasmon}
\end{equation}
where $n^\epsilon_c$ is the total energy defined in Table \ref{tab2} and $P = n^\epsilon_c/2$ is the pressure.

\begin{figure}
	\centering
	\includegraphics[width=0.3\textwidth]{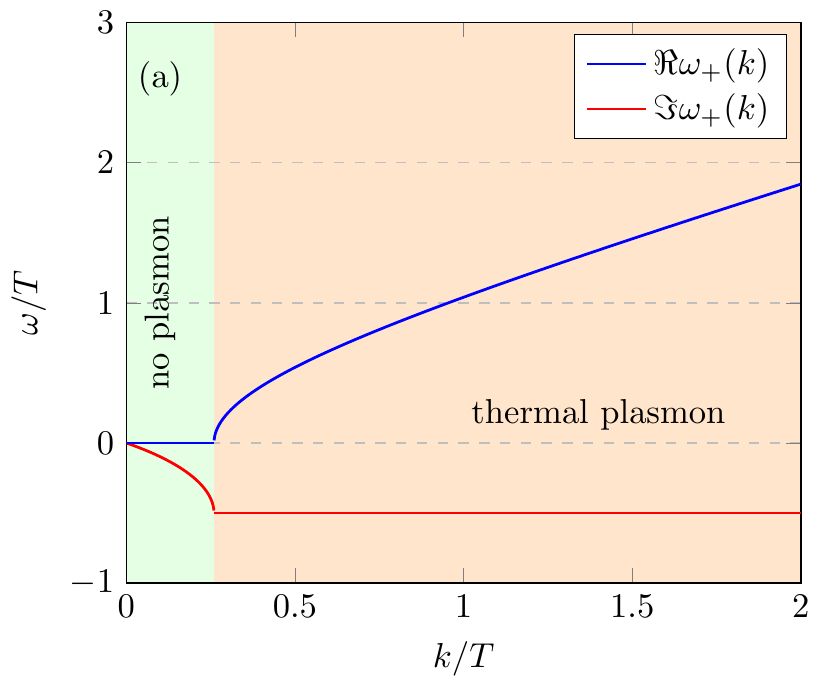}
	\includegraphics[width=0.3\textwidth]{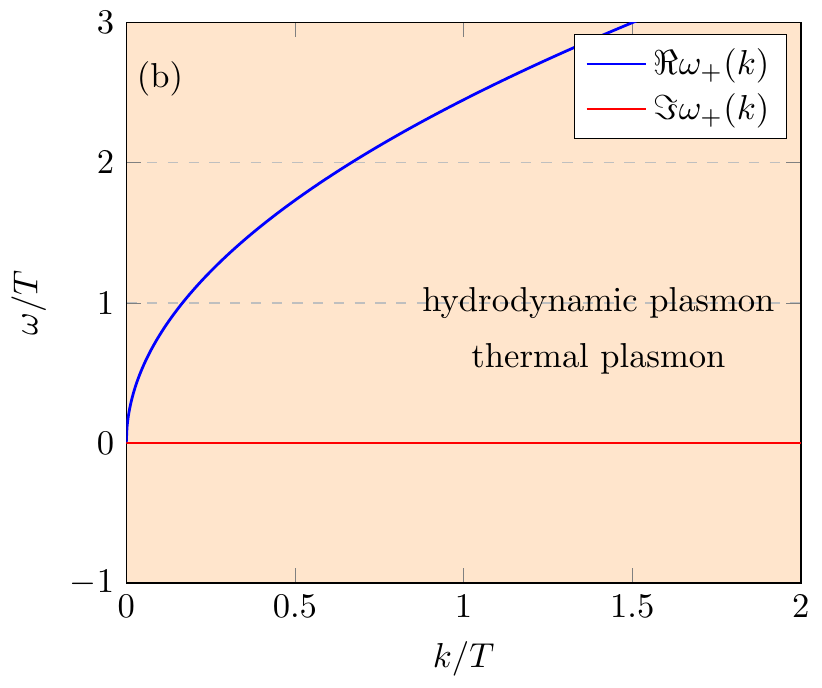}
	\caption{The plasmon energy dispersion in Eq.(\ref{eq:plasmonspectrum}) at two different chemical potentials: (a) at the charge neutrality point $\mu/T=0$ and (b) at $\mu/T = 10$. The plots show the plasmon dispersion in the physically distinct situations of $n_c=0$ and $n_c\neq0$. In the former case, only a thermal plasmon exists, whereas in the later a true hydrodynamic plasmon emerges. The frequency that separates the two regimes is related to the inverse drag time scale.}
	\label{fig:plasmonregimes}
\end{figure}

Fig.~\ref{fig:plasmonregimes} shows the plasmon dispersion in the two physically distinct situations: (i) at the charge neutrality point where $n_c = 0$ and (ii) away from the charge neutrality point where $n_c \neq 0$.

Note that the true hydrodynamic limit of the plasmons is obtained in the limit where $\tau_c$ and $\tau_{\text{imb}}$ are small. In that limit, the two components act as one fluid with the effective charge density $n_c$, and the individual components do not play a role. 
The corresponding plasmon frequency is shown in Fig.~\ref{fig:plasmonregimes} (b).
The important feature of Eq.~\eqref{eq:hydroplasmon} is that the plasmons cease to be excitations at the charge neutrality point where $n_c=0$, in agreement with hydrodynamic theory explained in Refs.~\cite{Lucas2018,Narozhny2019}, see Fig.~\ref{fig:plasmonregimes} (a). The RPA limit is obtained when the two components, electrons and holes, act independently. They conserve momentum, even on the individual level.
An important feature of this expression is that there remains a plasmon even at the charge neutrality point, $n_c=0$, see Fig.~\ref{fig:plasmonregimes} (a) and Eq.~\eqref{eq:plasmonspectrum}. It remains nonzero at the charge neutrality point if $T>0$.

\subsubsection{Discussion}
	In this section we have discussed some of the properties of two-component hydrodynamics of a fluid composed of electrons and holes which stem from two bands in an electronic lattice system. This situation has been considered before in multiple settings; including graphene or, more generically, Dirac systems and bilayer graphene. We have given special attention to drag terms coming from the collision integral that couples electrons and holes in the setup. From a methodological perspective, this is beyond pure hydrodynamics, which would be completely collision free. These effects are known to be required to determine the conductivity of such a system at the charge neutrality point, as demonstrated in graphene and bilayer graphene. We also set up the full thermo-electric response in this setup and recovered well-known literature results. 
	In a last part we investigated the role of collective modes. We especially considered the plasmon spectrum and establish a new result. Namely, we find that the same drag effects that are required to ensure a finite conductivity at charge neutrality govern a crossover from an RPA-type plasmon (that allows for a thermal plasmon) to the pure hydrodynamic plasmon that is not thermal. 

	\section{Part B: Electron-Hole-Plasmon theory}
	 It is known that in a Coulomb-interacting electronic system there exists a collective mode associated with plasma oscillations, called the plasmon~\cite{Bohm&Pines1953}. In typical three-dimensional metals, however, plasmons have a massive excitation gap. The size of the gap is typically larger than the Fermi energy $E_F$ of the underlying electronic system~\cite{Pines1953,Pines1999}. Consequently, for many practical questions, they only make negligible contributions to thermodynamic and transport properties. In contrast, plasmons in two dimensions are massless with a square-root type dispersion relation, {\it i.e.}, $\omega_{\text{pl}}(\vec{q}) \propto \sqrt{q}$~\cite{Stern1967,Wunsch2006,Sarma2013}. We showed above that this result can be reproduced from the hydrodynamic approach. One immediate consequence of the dispersion relation is that plasmons can easily be excited at experimentally accessible temperatures and therefore make a direct contribution to their transport properties; especially heat transport~\cite{Kitinan2020}. Another important feature is that the plasmons are remarkably stable~\cite{DasSarma2007}. To summarize, it has been shown that two-dimensional electronic systems host stable plasmons that are bona-fide quasi-particles on the quantum level. For a more technical discussion see our parallel paper.

\subsection{The setup}\label{subsec:setup}		 
In this part of the paper we discuss the phenomenological set of coupled Boltzmann equations introduced in Eq.~\eqref{eq:coupledBoltzmannfull}. This describes a system that consists of electrons, holes, and plasmons and treats them on equal footing. With minor modifications, this formalism can also account for other collective modes like phonons with perfect drag, spin-waves or similar degrees of freedom by modifying the dispersion relation.
		 
	 In a companion paper, we present a formal derivation of the Boltzmann equations. It starts from an effective field theory for plasmons emerging from underlying two-dimensional Coulomb-interacting Dirac electrons. That work is based on real-time quantum field theory along a closed time-contour, typically referred to as Keldysh quantum field theory. From a technical point of view, the key insight of that paper is a formal derivation of the Boltzmann equations, where we show explicitly that all conservation laws hold which here we just assume.

The coupled Boltzmann equations read
\begin{widetext}
\begin{eqnarray}\label{eq:electronplasmonBoltzmann}
		\partial_t f_++ \partial_{\vec{k}} \epsilon_+(\vec{k},\vec{x}, t) \cdot \partial_{\vec{x}} f_+- \partial_{\vec{x}} \epsilon_+ (\vec{k},\vec{x}, t)\cdot\partial_{\vec{k}} f_+&=&\mathcal{I}_+[f_+,f_-,b] \nonumber \\
		\partial_t f_-+ \partial_{\vec{k}} \epsilon_-(\vec{k},\vec{x}, t) \cdot \partial_{\vec{x}} f_- -\partial_{\vec{x}} \epsilon_- (\vec{k},\vec{x}, t)\cdot\partial_{\vec{k}} f_-&=&\mathcal{I}_{-}[f_-,f_+,b]\nonumber \\  \partial_t b+ \partial_{\vec{k}} \omega(\vec{k},\vec{x}, t) \cdot \partial_{\vec{x}} b -\partial_{\vec{x}} \omega (\vec{k},\vec{x}, t)\cdot\partial_{\vec{k}} b&=&\mathcal{I}_{b}[b,f_+,f_-]  \;.\nonumber \\
	\end{eqnarray}	
\end{widetext} 
	where $\mathcal{I}_+[f_+,f_-,b]$, $\mathcal{I}_{-}[f_-,f_+,b]$, and $\mathcal{I}_{b}[b,f_+,f_-]$ contain all the terms introduced in Eq.~\eqref{eq:coupledBoltzmannfull}.

In order to make progress, we again set up a set of three linearized coupled Boltzmann equations within the relaxation-time approximation. We make one important assumption here which is that the plasmon sector cannot equilibrate by itself. Instead, it is coupled to the fermion sector via perfect drag and all the total momentum relaxation happens in the fermionic sector. In that sense this scenario is very similar to a scenario with perfect electron-phonon drag.
	We showed in the previous section that the plasmon energy dispersion depends on fermion density, see Eqs.~\eqref{eq:hydroplasmon} and~\eqref{eq:plasmonspectrum}, which in turn is space and time dependent. As a result, there is a density-gradient force, -$\partial_{\vec{x}} \omega(\vec{k},\vec{x}, t)$, exerted on the plasmons due to a difference in fermion density across a volume element. In addition, here the fermion energy assumes space-dependence leading to a conservative force term, -$\partial_{\vec{x}}\epsilon_{\pm}(\vec{k},\vec{x}, t)$, in the  Boltzmann equations. This is the collective force due to the other particles in the system, for example, the local Hartree and Fock potentials.

\subsection{Thermo-electric response}\label{subsec:thermoelectricII}

In order to make progress, we again set up a set of three linearized coupled Boltzmann equations within the relaxation-time approximation.
	
\begin{figure}[tbh!]
	\centering	\includegraphics[scale=0.7]{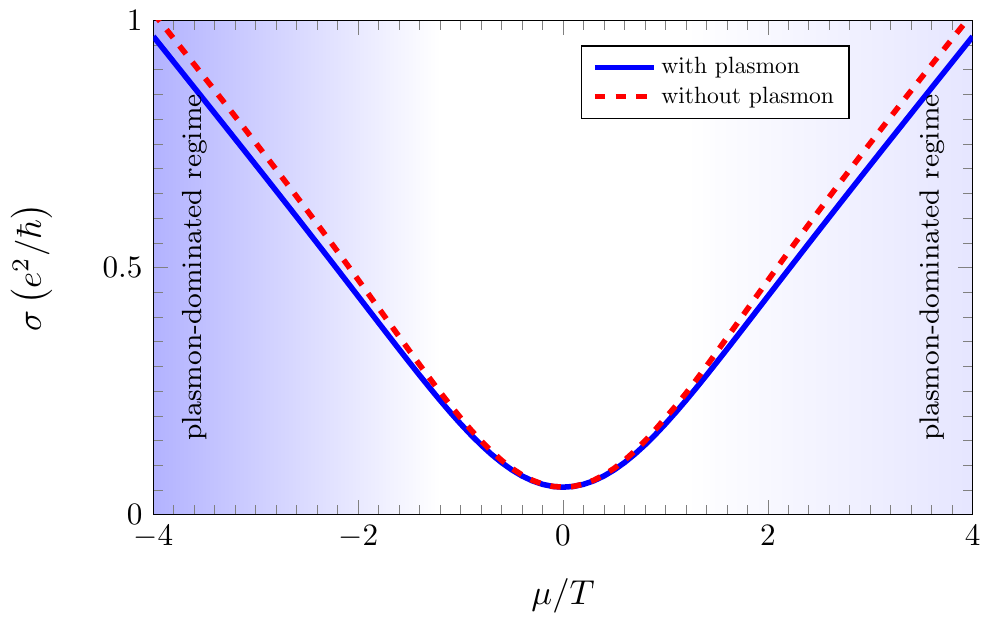}
	\includegraphics[scale=0.7]{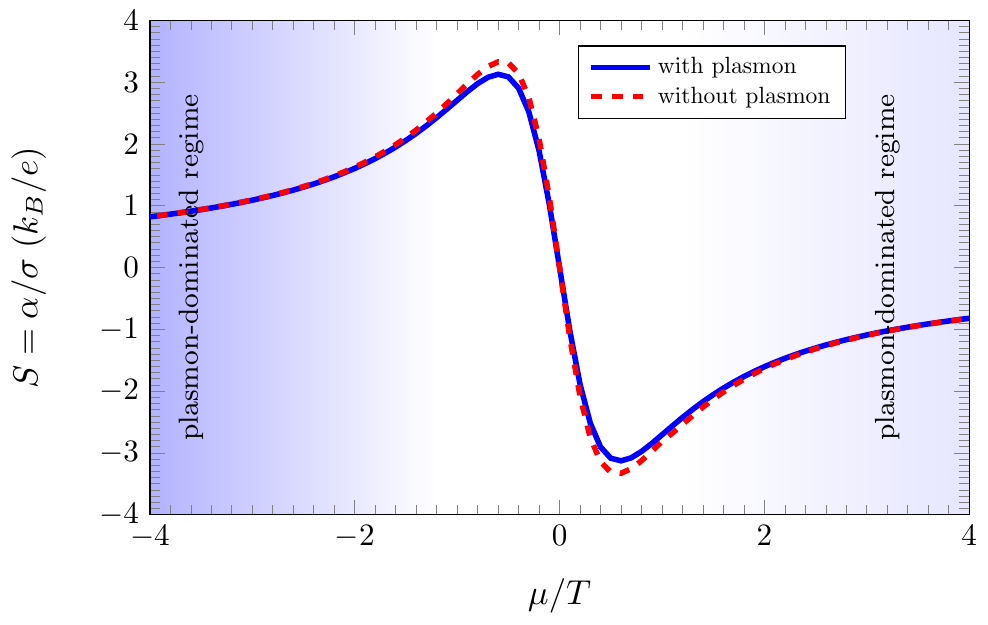}
	\caption{Upper panel: The conductivity calculated within the pure electron-hole theory and within the electron-hole-plasmon theory. Lower panel: The same comparison for the Seebeck coefficient.}
	\label{fig:transportplasmon}
\end{figure}

\begin{figure}[tbh!]
	\includegraphics[scale=0.7]{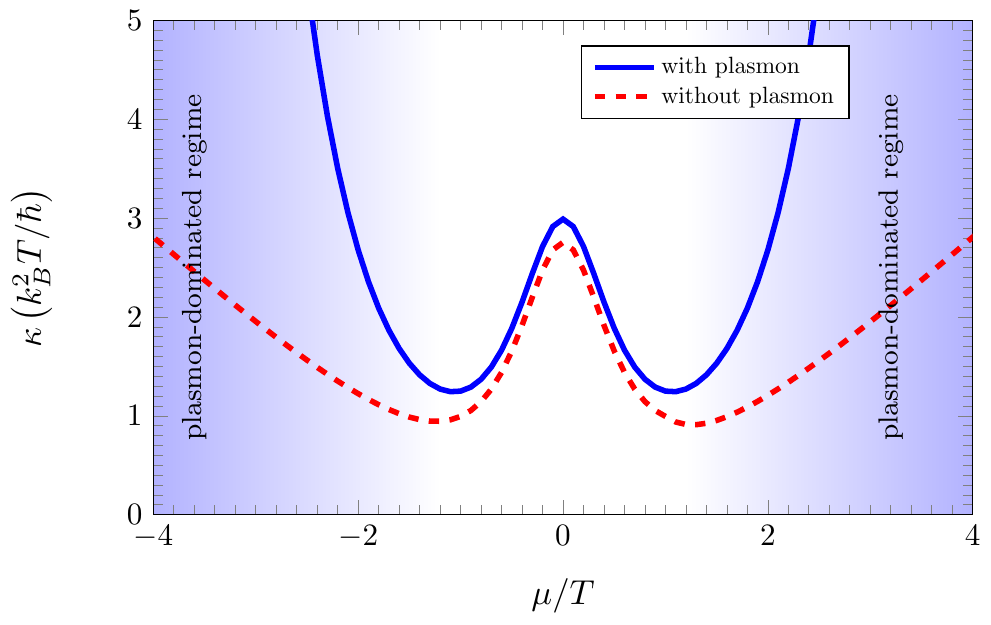}
	\includegraphics[scale=0.7]{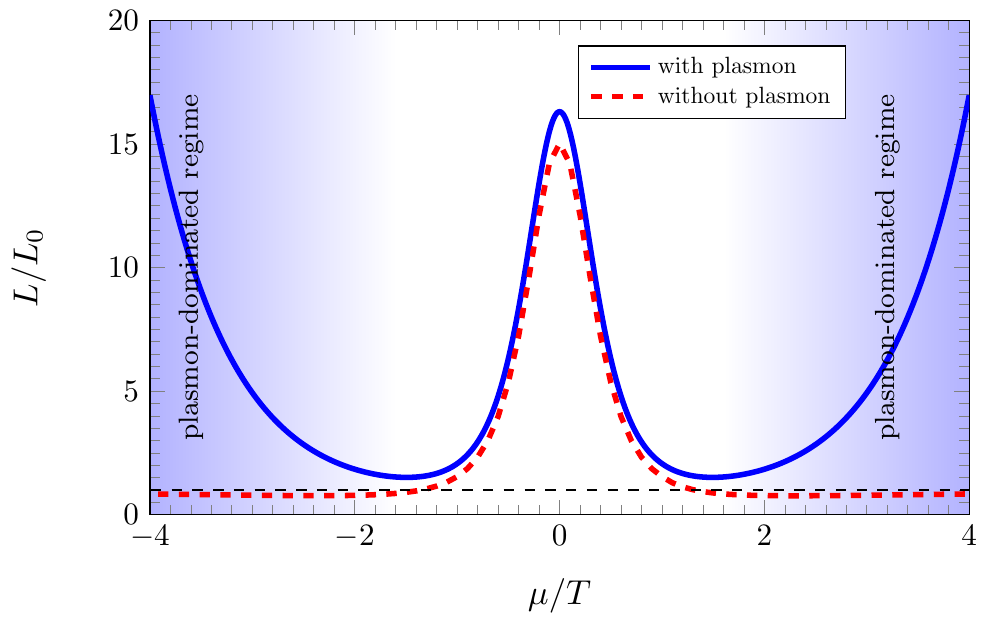}
	\caption{ Upper panel: The heat conductivity without and with the plasmon contribution. Lower panel: The same comparison for the Lorenz ratio.}
	\label{fig:Wiedemann-Franzratio}
\end{figure}

	\begin{widetext}
\begin{eqnarray}
\partial_t \delta f_+-\frac{\epsilon_+-\mu}{T}  \partial_{\vec{x}}T\cdot \partial_{\vec{k}}f^0_+  - e \vec{E}\cdot \partial_{\vec{k}}f^0_+&=& -\frac{\delta f_+}{\tau_{+-}}+\frac{\delta f_-}{\tau_{-+}} -\frac{\delta f_+}{\tau^{\rm{dis}}_+}-\frac{\delta f_+}{\tau_{+b}}+\frac{\delta b}{\tau_{b+}}\;, \nonumber \\
\partial_t \delta f_--\frac{\epsilon_--\mu}{T}  \partial_{\vec{x}}T\cdot \partial_{\vec{k}}f^0_- - e \vec{E}\cdot \partial_{\vec{k}}f^0_-&=& -\frac{\delta f_-}{\tau_{-+}}+\frac{\delta f_+}{\tau_{+-}} -\frac{\delta f_-}{\tau^{\rm{dis}}_-}-\frac{\delta f_-}{\tau_{-b}}+\frac{\delta b}{\tau_{b-}}  \;,\nonumber \\ \partial_t \delta b- \frac{\omega}{T}\partial_{\vec{x}}T \cdot \partial_{\vec{k}}b^0&=&-\delta b \left(\frac{1}{\tau_{b+}}+\frac{1}{\tau_{b-}} \right)+\frac{\delta f_+}{\tau_{+b}}+\frac{\delta f_-}{\tau_{-b}}\;.
\end{eqnarray}
\end{widetext}
As in the previous discussion, we choose the relaxation times such that they obey the conservation laws of charge, momentum (this is broken in presence of disorder), and energy.
The equations can be solved for any given realization of the scattering times but generally lead to bulky expressions. The charge and heat currents read
\begin{widetext}
\begin{eqnarray}
\vec{j}_c=-e\int_{\vec{k}} \partial_{\vec{k}} \epsilon_+ \left(\delta f_+-\delta f_- \right) \quad {\rm{and}} \quad \vec{j}_Q=\int_{\vec{k}} \partial_{\vec{k}} \epsilon_+  \left(\left(\epsilon_+-\mu \right)\delta f_++ \left(\epsilon_++\mu \right) \delta f_- \right) +\int_{\vec{k}} \partial_{\vec{k}} \omega\, \omega\, \delta b\;,
\end{eqnarray}
\end{widetext}
where we again used $\epsilon_+=-\epsilon_-$. In the following we only discuss the d.c. properties ($\partial_t \delta f_+=\partial_t \delta f_-=\partial_t \delta b=0$) at the charge neutrality point, $\mu=0$, which leads to a number of simplifications. The first one concerns the relaxation times. At charge neutrality, we have $\tau_{+-}=\tau_{-+}$, $\tau_+^{\rm{dis}}=\tau_-^{\rm{dis}}$, $\tau_{+b}=\tau_{-b}$, and $\tau_{b+}=\tau_{b-}$. It is furthermore straightforward to show that under those conditions there is no drag on the plasmon sector from an applied electric field, meaning $\delta b = 0$. The reason is very simple: an electric field only couples to the electrons and holes. They are accelerated into different directions leading to a zero momentum state. Consequently, there is no momentum transfer into the plasmon sector, meaning the plasmons remain in equilibrium. This argument can also be generalized to the off-diagonals of the conductivity tensor, which was also true in the case without plasmons. In the hydrodynamic limit, we are interested in the limit $1/\tau_{+}^{\rm{dis}} \ll 1/\tau_{+b}$, valid for very clean systems. Consequently, we find
\begin{widetext}
\begin{eqnarray}
\sigma=e^2  \frac{\tau_{+b}\tau_+^{\rm{dis}}}{2 \tau_{+b}+\tau_+^{\rm{dis}}}  \int_{\vec{k}}\partial_{\vec{k}} \epsilon_+ \cdot \partial_{\vec{k}} \left(f_+^0-f_-^0 \right) \approx e^2 \tau_{+b}\int_{\vec{k}}\partial_{\vec{k}} \epsilon_+ \cdot \partial_{\vec{k}} \left(f_+^0-f_-^0 \right)
\end{eqnarray}
for the electrical conductivity and
\begin{eqnarray}
\kappa&=& \frac{\tau_{+}^{\rm{dis}}}{T} \int_{\vec{k}} \left( \epsilon_+ \partial_{\vec{k}}\left( f_+^0-f_-^0     \right)+ \omega \partial_{\vec{k}}b^0 \right) \cdot \left(\epsilon_+ \partial_{\vec{k}} \epsilon_+ + \frac{1}{2}\frac{\tau_{b+}}{\tau_{+b}} \omega \partial_{\vec{k}} \omega \right) + \frac{\tau_{b+}}{2T} \int_{\vec{k}} \partial_{\vec{k}} \omega \cdot \partial_{\vec{k}}b^0  \omega^2  \nonumber \\ &\approx&  \frac{\tau_{+}^{\rm{dis}}}{T} \int_{\vec{k}} \left( \epsilon_+ \partial_{\vec{k}}\left( f_+^0-f_-^0     \right)+ \omega \partial_{\vec{k}}b^0 \right) \cdot \left(\epsilon_+ \partial_{\vec{k}} \epsilon_+ + \frac{1}{2}\frac{\tau_{b+}}{\tau_{+b}} \omega \partial_{\vec{k}} \omega \right)\;
\end{eqnarray}
for the thermal conductivity.
It is important to point out that, again, the charge conductivity is finite through drag-scattering with the plasmon sector, even in the clean limit. The heat conductivity, on the other hand, diverges, as we expect. Furthermore, we find that the heat conductivity receives a contribution from the plasmon sector which is, a priori, on the same order of magnitude as the electronic contribution.
\end{widetext}
Away from the charge neutrality point, there will again be a Drude peak that signals diverging electrical conductivity in the absence of disorder. Since the corresponding expressions are very bulky, we refrain from showing their explicit form. However, we present a plot of the most important thermo-electric coefficients in  Figs.~\ref{fig:transportplasmon} and~\ref{fig:Wiedemann-Franzratio}. In Fig.~\ref{fig:transportplasmon}, we show the electrical conductivity, as well as the Seebeck coefficient. Importantly, we find excellent agreement between the theory of only electrons discussed in Sec.~\ref{subsec:thermoelectricI} and Fig.~\ref{fig:sigma}, and the one of electrons, holes, and plasmons drag coupled. However, looking at Fig.~\ref{fig:Wiedemann-Franzratio}, we find a strong enhancement of the thermal conductivity, that grows upon tuning away from charge neutrality. For this plot, we have used the same parameters as for Fig.~\ref{fig:transportplasmon}. This is in agreement with a more sophisticated solution of the transport equations that was presented in Ref.~\cite{Kitinan2020} by some of the authors.

\section{Conclusion and Outlook}\label{sec:conclusion}
In this paper, we investigated the full thermo-electric transport properties of an interacting fluid of electrons, holes, and a collective mode from a theoretical point of view. In the hydrodynamic limit we used an approach based on a Boltzmann transport equation within a relaxation-time approximation that mimics a full numerical solution of the set of equations. We treated electrons, holes and plasmons on  equal footing and discussed the hydrodynamic behavior of the combined system. This includes conservation laws but, in particular, we discussed the thermo-electric behavior in the very clean system. One of our findings is that in that situation the plasmons make a sizeable contribution to transport properties and, therefore, cannot be neglected. We hypothesize that this also applies to other quantities sensitive to the energy-momentum tensor, such as the viscosity. The approach we develop can also be applied in a straightforward manner to any hybrid system of electrons, holes, and bosons. In a parallel paper we show the technical underpinnings of our theory based on the Schwinger-Keldysh approach. Not only do we derive the set of coupled Boltzmann equations there, we also establish explicitly all the conservation laws that we use here.

\section*{Acknowledgements}
 We acknowledge former collaborations and discussions with R.A. Duine, M. Fremling, P.M. Gunnink, J.S. Harms, E.I. Kiselev, A. Lucas, J. Lux, A. Mitchell, M. M\"uller, A. Rosch, S. Sachdev, J. Schmalian, M. Vojta, G. Wagner, and J. Waissman. KP thanks the Institute for the Promotion of Teaching Science and Technology (IPST) of Thailand for a Ph.D. fellowship. TL thanks the Dutch Research Council (NWO) for (partly) financing this work as a part of the research programme Fluid Spintronics with project number 182.069. This work is part of the D-ITP consortium, a program of the Netherlands Organisation for Scientific Research (NWO) that is funded by the Dutch Ministry of Education, Culture and Science
	(OCW). 

	\pagebreak
	\clearpage
	\newpage
	\widetext
	\appendix
\begin{center}
	\textbf{\large Appendices}
\end{center}
\section{Numerical evaluation of the scattering time for electron-hole drag}\label{sec:scatteringtime}
In this appendix, we discuss how to perform the relaxation-time approximation for electron-hole drag on a systematic level . To this end, we consider a set of coupled Boltzmann equations for a two-dimensional Dirac electron-hole plasma. In the companion paper, we derive these equations from a Keldysh quantum field theory within a perturbation theory up to second order. The equations read
\begin{eqnarray}
	&&\partial_t f_\lambda(\vec{x},t,\vec{k}) +\lambda v_F \hat{k}\cdot \partial_{\vec{x}} f_\lambda(\vec{x},t,\vec{k})-\; e\vec{E}\cdot \partial_{\vec{k}} f_\lambda(\vec{x},t,\vec{k})
	\nonumber\\&&	\hspace{3cm}= -\int \frac{d\vec{k}_1}{(2\pi)^2}\frac{d\vec{q}}{(2\pi)^2} 2\pi\delta(\epsilon_{\lambda}(\vec{k})-\epsilon_{\lambda_1}(\vec{k}-\vec{q})-\epsilon_{\lambda_2}(\vec{k}_1+\vec{q})+\epsilon_{\lambda_3}(\vec{k}_1))R_{\lambda\lambda_1\lambda_3\lambda_2}(\vec{k},\vec{k}_1,\vec{q})\nonumber\\&&\hspace{3cm}\Big[f_{\lambda}(\vec{k})f_{\lambda_3}(\vec{k}_1)(1-f_{\lambda_1}(\vec{k}-\vec{q}))(1-f_{\lambda_2}(\vec{k}_1+\vec{q}))-(1-f_{\lambda}(\vec{k}))(1-f_{\lambda_3}(\vec{k}_1))f_{\lambda_1}(\vec{k}-\vec{q})f_{\lambda_2}(\vec{k}_1+\vec{q})\Big].\nonumber\\
	\label{eq:fermionboltzmannequation}
\end{eqnarray}
We introduce the shorthand notation for the transition probability that is
\begin{equation}
	R_{\lambda\lambda_1\lambda_3\lambda_2}(\vec{k},\vec{k}_1,\vec{q})=2\Big[|T_{\lambda\lambda_1\lambda_3\lambda_2}-T_{\lambda\lambda_2\lambda_1\lambda_3}|^2+(N-1)\left(|T_{\lambda\lambda_1\lambda_3\lambda_2}|^2+|T_{\lambda\lambda_2\lambda_1\lambda_3}|^2\right) \Big],
\end{equation}
where
\begin{equation}
	T_{\lambda \lambda_1\lambda_2\lambda_3}(\vec{k},\vec{k}_1,\vec{q}) = \frac{V(\vec{q})}{2} M^{\lambda\lambda_1}_{\vec{k},\vec{k}-\vec{q}}M^{\lambda_2\lambda_3}_{\vec{k}_1,\vec{k}_1+\vec{q}}\;
\end{equation}
with $N$ being the number of flavors.
The Coulomb interaction in Fourier space is $V(\vec{q}) = 2\pi \alpha/q$, where $\alpha$ is the fine-structure constant characterizing the strength of the interaction. The coherence factor coming from the overlap of the wavefunction is defined according to
\begin{equation}
	M^{\lambda\lambda_1}_{\vec{k},\vec{k}_1}=\left(\mathcal{U}^\dagger_{\vec{k}}\mathcal{U}_{\vec{k}_1}\right)_{\lambda\lambda_1}\;,
\end{equation}
and
\begin{equation}
	\mathcal{U}_{\vec{k}} = \frac{1}{\sqrt{2}}\begin{pmatrix}
		-\exp(-i\theta_{\vec{k}})&\exp(-i\theta_{\vec{k}})\\1&1
	\end{pmatrix}.
	\label{eq:tranformationmatrix}
\end{equation} 
Here $\tan(\theta_{\vec{k}})=k_y/k_x$ and $k$ denotes the magnitude of the momentum $\vec{k}$.
Moreover, $\lambda$, $\lambda_1$, $\lambda_2$, $\lambda_3 = \pm$ are energy-band indices, $+$\; for the conduction band and $-$ \;for the valence band.  The collision integral on the right-hand side describes a process that an electron in the energy band $\lambda$ and the momentum state $\vec{k}$  is scattered into the energy band $\lambda_1$ and momentum state $\vec{k}+\vec{q}$ by a collision with another electron in band $\lambda_3$ and state $\vec{k}_1$, which is itself scattered into the energy band $\lambda_2$ and state $\vec{k}_1-\vec{q}$. The initial states $\vec{k}$ and $\vec{k}_1$ have to be filled and the final states $\vec{k}-\vec{q}$ and $\vec{k}_1+\vec{q}$ must be empty for this event to take place. The factors $f_{\lambda}(\vec{k})$ and $f_{\lambda_3}(\vec{k}_1)$ are the occupation number of these state. $1-f_{\lambda_1}(\vec{k}-\vec{q})$ and $1-f_{\lambda_2}(\vec{k}_1+\vec{q})$ describes the probability that the final states are unoccupied. The conservation of energy is taking into account by the delta function. The transition probability of this event is
$R_{\lambda\lambda_1\lambda_3\lambda_2}$. 

\begin{figure}
	\centering
	\includegraphics[width=0.4\textwidth]{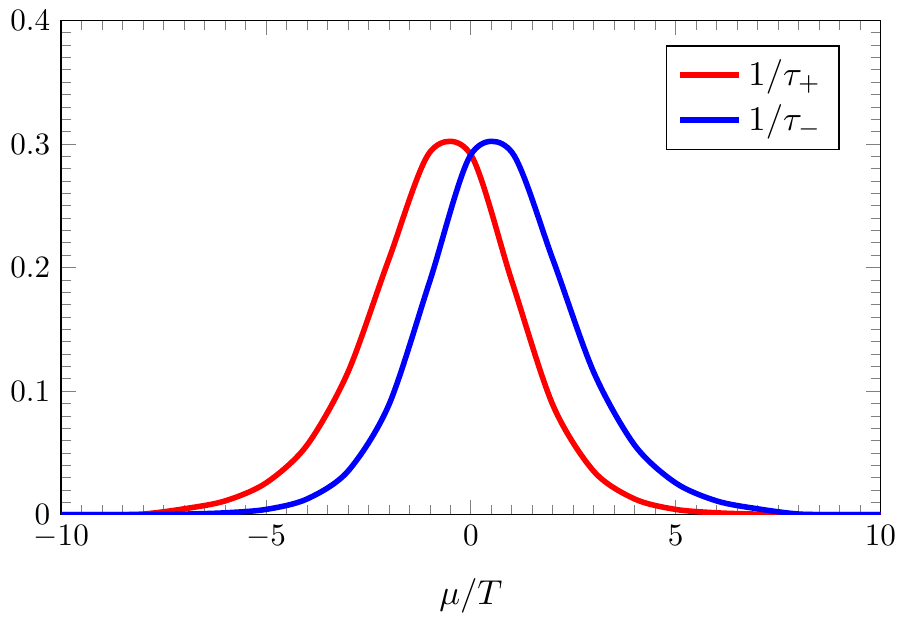}
	\caption{Numerical solution of the relaxation time for electron-hole drag as a function of chemical potential $\mu$.}
	\label{fig:taudragnumerics}
\end{figure}

Next, we consider the electron-hole drag part of the collision integral that is when
$\lambda=\lambda_1$ and $\lambda_2=\lambda_3=-\lambda$\;. Moreover, we are interested in a steady state and homogeneous solution, so the time-derivative and space-derivative term on the left-hand side of the Boltzmann equations are zero. In the linear-response regime, we may solve the Boltzmann-equations to linear order using the ansatz  $f_{\lambda}(\vec{k}) = f^0_\lambda(\vec{k})+\delta f_\lambda(\vec{k})$, where
\begin{equation}
	\delta f_\lambda(\vec{k})=g_{\lambda}(\vec{k})e\vec{E}\cdot \partial_{\vec{k}}f^0_\lambda(\vec{k})=g_{\lambda}(\vec{k})e\vec{E}\cdot \lambda v_F \frac{\vec{k}}{k}f^0_\lambda(\vec{k})(1-f_{\lambda}^0(\vec{k}))
\end{equation}
The unknown function $g_{\lambda}(k)$ is determined by solving the Boltzmann equations. It is usually expanded in terms of an appropriate set of basis functions
\begin{equation}
	g_\lambda(\vec{k}) = \sum_{n,p} \tau^{(n,p)}_{\lambda}\phi^{(n,p)}_{\lambda}(k),
\end{equation}
where $\tau^{(n,p)}_\lambda$ can be interpreted as a relaxation time for the corresponding mode.
Here, we use
\begin{equation}
	\phi^{(n,p)}_{\lambda}(k)= \lambda^p k^n.
\end{equation}
After linearizing the Boltzmann equations, we obtain
\begin{eqnarray}
	\partial_{\vec{k}}f^0_{\lambda}(\vec{k})&=&-\int \frac{d\vec{k}_1}{(2\pi)^2}\frac{d\vec{q}}{(2\pi)^2}\mathcal{P}(\vec{k},\vec{k}_1,\vec{q})\Big[f^0_{\lambda}(\vec{k})f^0_{-\lambda}(\vec{k}_1)(1-f^0_{\lambda}(\vec{k}-\vec{q}))(1-f^0_{-\lambda}(\vec{k}_1+\vec{q}))\Big]\nonumber\\&&\Big[\lambda v_F \frac{\vec{k}}{k}g_{\lambda}(k) -\lambda v_F \frac{\vec{k}_1}{k_1}g_{-\lambda}(k_1)-\lambda v_F \frac{\vec{k}-\vec{q}}{|\vec{k}-\vec{q}|}g_{\lambda}(|\vec{k}-\vec{q}|)+\lambda v_F \frac{\vec{k}_1+\vec{q}}{|\vec{k}_1+\vec{q}|}g_{-\lambda}(|\vec{k}_1+\vec{q}|)  \Big].
\end{eqnarray}
Here we introduce a short-hand notation
\begin{equation}
	\mathcal{P}(\vec{k},\vec{k}_1,\vec{q}) = 2\pi\delta(\epsilon_{\lambda}(\vec{k})-\epsilon_{\lambda_1}(\vec{k}-\vec{q})-\epsilon_{\lambda_2}(\vec{k}_1+\vec{q})+\epsilon_{\lambda_3}(\vec{k}_1))R_{\lambda\lambda_1\lambda_3\lambda_2}(\vec{k},\vec{k}_1,\vec{q}).
\end{equation}
We are particularly interested in current relaxation, so we look at the current mode $\phi_\lambda^{(0,1)}(k)=\lambda$ and assume that the other modes are not excited. By projecting the Boltzmann-equation onto this mode, we find
\begin{eqnarray}
	\int\frac{d\vec{k}}{(2\pi)^2}\phi_\lambda^{(0,1)}(k)\frac{\vec{k}}{k}\cdot\partial_{\vec{k}}f^0_{\lambda}(\vec{k})&=&-\int\frac{d\vec{k}}{(2\pi)^2} \frac{d\vec{k}_1}{(2\pi)^2}\frac{d\vec{q}}{(2\pi)^2}\mathcal{P}(\vec{k},\vec{k}_1,\vec{q})\Big[f^0_{\lambda}(\vec{k})f^0_{-\lambda}(\vec{k}_1)(1-f^0_{\lambda}(\vec{k}-\vec{q}))(1-f^0_{-\lambda}(\vec{k}_1+\vec{q}))\Big]\nonumber\\&&\phi_{\lambda}^{(0,1)}(k)\frac{\vec{k}}{k}\cdot\Big[\lambda v_F \frac{\vec{k}}{k}\tau_{\lambda}^{(0,1)} \phi^{(0,1)}_{\lambda}(k) -\lambda v_F \frac{\vec{k}_1}{k_1}\tau_{-\lambda}^{(0,1)} \phi^{(0,1)}_{-\lambda}(k_1)\nonumber\\&&-\lambda v_F \frac{\vec{k}-\vec{q}}{|\vec{k}-\vec{q}|}\tau_{\lambda}^{(0,1)} \phi^{(0,1)}_{\lambda}(|\vec{k}-\vec{q}|)+\lambda v_F \frac{\vec{k}_1+\vec{q}}{|\vec{k}_1+\vec{q}|}\tau_{-\lambda}^{(0,1)} \phi^{(0,1)}_{-\lambda}(|\vec{k}_1+\vec{q}|)  \Big].
\end{eqnarray}
To evaluate the current relaxation time $\tau_\lambda^{(0,1)}$, we again assume that the mode corresponding to $\phi_{-\lambda}^{(0,1)}$ is not excited, this gives
\begin{eqnarray}
	&&\frac{1}{\tau^{(0,1)}_\lambda}\times\left(\int\frac{d\vec{k}}{(2\pi)^2}\phi_\lambda^{(0,1)}(k)\frac{\vec{k}}{k}\cdot\partial_{\vec{k}}f^0_{\lambda}(\vec{k})\right) =\nonumber\\&&\hspace{1cm} -\int\frac{d\vec{k}}{(2\pi)^2} \frac{d\vec{k}_1}{(2\pi)^2}\frac{d\vec{q}}{(2\pi)^2}\mathcal{P}(\vec{k},\vec{k}_1,\vec{q})\Big[f^0_{\lambda}(\vec{k})f^0_{-\lambda}(\vec{k}_1)(1-f^0_{\lambda}(\vec{k}-\vec{q}))(1-f^0_{-\lambda}(\vec{k}_1+\vec{q}))\Big]\nonumber\\&&\hspace{1cm}\phi_{\lambda}^{(0,1)}(k)\frac{\vec{k}}{k}\cdot\Big[\lambda v_F \frac{\vec{k}}{k} \phi^{(0,1)}_{\lambda}(k) -\lambda v_F \frac{\vec{k}-\vec{q}}{|\vec{k}-\vec{q}|} \phi^{(0,1)}_{\lambda}(|\vec{k}-\vec{q}|) \Big],
\end{eqnarray}
and similarly $\tau_{-\lambda}^{(0,1)}$ is determined from
\begin{eqnarray}
	&&\frac{1}{\tau^{(0,1)}_{-\lambda}}\times\left(\int\frac{d\vec{k}}{(2\pi)^2}\phi_\lambda^{(0,1)}(k)\frac{\vec{k}}{k}\cdot\partial_{\vec{k}}f^0_{\lambda}(\vec{k})\right) =\nonumber\\&&\hspace{1cm} -\int\frac{d\vec{k}}{(2\pi)^2} \frac{d\vec{k}_1}{(2\pi)^2}\frac{d\vec{q}}{(2\pi)^2}\mathcal{P}(\vec{k},\vec{k}_1,\vec{q})\Big[f^0_{\lambda}(\vec{k})f^0_{-\lambda}(\vec{k}_1)(1-f^0_{\lambda}(\vec{k}-\vec{q}))(1-f^0_{-\lambda}(\vec{k}_1+\vec{q}))\Big]\nonumber\\&&\hspace{1cm}\phi_{\lambda}^{(0,1)}(k)\frac{\vec{k}}{k}\cdot\Big[-\lambda v_F \frac{\vec{k}_1}{k_1} \phi^{(0,1)}_{-\lambda}(k_1) +\lambda v_F \frac{\vec{k}_1+\vec{q}}{|\vec{k}_1+\vec{q}|} \phi^{(0,1)}_{-\lambda}(|\vec{k}_1+\vec{q}|) \Big].
\end{eqnarray}
These integrals can be evaluate numerically. In Fig.~\ref{fig:taudragnumerics}, we plot the relaxation time for electron-hole drag  $1/\tau_{\pm}$, which we defined from the discussion above according to
\begin{eqnarray}
	\frac{1}{\tau_+} &=&-\int\frac{d\vec{k}}{(2\pi)^2} \frac{d\vec{k}_1}{(2\pi)^2}\frac{d\vec{q}}{(2\pi)^2}\mathcal{P}(\vec{k},\vec{k}_1,\vec{q})\Big[f^0_{+}(\vec{k})f^0_{-}(\vec{k}_1)(1-f^0_{+}(\vec{k}-\vec{q}))(1-f^0_{-}(\vec{k}_1+\vec{q}))\Big]\nonumber\\&&\phi_{+}^{(0,1)}(k)\frac{\vec{k}}{k}\cdot\Big[ v_F \frac{\vec{k}}{k} \phi^{(0,1)}_{+}(k) - v_F \frac{\vec{k}-\vec{q}}{|\vec{k}-\vec{q}|} \phi^{(0,1)}_{+}(|\vec{k}-\vec{q}|) \Big]\;,\nonumber\\
	\frac{1}{\tau_-} &=& -\int\frac{d\vec{k}}{(2\pi)^2} \frac{d\vec{k}_1}{(2\pi)^2}\frac{d\vec{q}}{(2\pi)^2}\mathcal{P}(\vec{k},\vec{k}_1,\vec{q})\Big[f^0_{+}(\vec{k})f^0_{-}(\vec{k}_1)(1-f^0_{+}(\vec{k}-\vec{q}))(1-f^0_{-}(\vec{k}_1+\vec{q}))\Big]\nonumber\\&&\hspace{1cm}\phi_{+}^{(0,1)}(k)\frac{\vec{k}}{k}\cdot\Big[- v_F \frac{\vec{k}_1}{k_1} \phi^{(0,1)}_{-}(k_1) + v_F \frac{\vec{k}_1+\vec{q}}{|\vec{k}_1+\vec{q}|} \phi^{(0,1)}_{-}(|\vec{k}_1+\vec{q}|) \Big]\;.
\end{eqnarray}	 
\noindent We find that the relaxation time decays exponentially in the large-$\mu$ limit which is the Fermi-liquid regime. One may understand this by observing that the relaxation time takes the form
\begin{eqnarray}
	\frac{1}{\tau_\pm} &=& \int f^0_{+}(\vec{k})f^0_{-}(\vec{k}_1)(1-f^0_{+}(\vec{k}-\vec{q}))(1-f^0_{-}(\vec{k}_1+\vec{q}))\Big[\dots\Big],\nonumber\\
	&=& \int \frac{1}{e^{\frac{k-\mu}{T}}+1}\frac{1}{e^{\frac{-k_1-\mu}{T}}+1}\frac{1}{e^{\frac{-|\vec{k}-\vec{q}|+\mu}{T}}+1}\frac{1}{e^{\frac{|\vec{k}_1+\vec{q}|+\mu}{T}}+1}\Big[\dots\Big].
\end{eqnarray}
In the large-$\mu$ limit, these become
\begin{equation}
	\frac{1}{\tau_\pm}
	\approx \int \Theta(\mu-k)\Theta(|\vec{k}-\vec{q}|-\mu) \exp(-\mu/T) \Big[\dots\Big].
\end{equation}
\section{Plasmon mode within the RPA approximation}
In this appendix, we derive the energy dispersion relation as well as the Landau damping for plasmon in a two-dimensional Dirac system within the RPA approximation. Here, we use units in which $v_F=1$. To this end, let us consider the RPA dielectric function 
\begin{equation}
	\epsilon_{\rm{RPA}}(\vec{q},\omega) = 1-V(\vec{q})\Pi(\vec{q},\omega)\;.
	\label{dielectricRPA}
\end{equation}	
The polarization function $\Pi(\vec{k},\omega)$ is given by the Lindhard formula
\begin{equation}	
	\Pi(\vec{q},\omega) =N\sum_{\lambda\lambda'=\pm1} \int\frac{d\vec{p}}{(2\pi)^2} \mathcal{F}_{\lambda\lambda'}(\vec{p},\vec{q})
	\frac{f^0_\lambda(\vec{p})-f^0_{\lambda'}(\vec{q}+\vec{p})}{\omega+i0^++\epsilon_\lambda (\vec{p})-\epsilon_{\lambda'}(\vec{q}+\vec{p})},
	\label{eq:polarizationfunction}
\end{equation}
where $\lambda,\lambda'=\pm$ denote the energy band; $+$ denotes the conduction band and $-$ denotes the valence band. The coherence factor is defined according to
\begin{equation} \mathcal{F}_{\lambda\lambda'}(\vec{p},\vec{q})= \frac{1}{2}(1+\lambda\lambda'\cos(\theta_{\vec{p}+\vec{q}}-\theta_{\vec{q}})).
\end{equation}
The Fermi-Dirac distribution function is
\begin{equation}
	f^0_\pm(\vec{p})=\frac{1}{e^{\frac{\epsilon_\pm (\vec{p})-\mu}{T}}+1}.
\end{equation}
Note that here we use a different notation from the main text and work in terms of electrons only instead of electrons and holes.
The plasmon frequency ($\omega=\omega_p-i\gamma_p$) is obtained from zero of the RPA dielectric function. Let us note that by defining it in this way the decay rate $\gamma_p$ is positive. If the damping is sufficiently weak ($\gamma_p \ll \omega_p$), we can expand the polarization function to leading order in $\gamma_p$. This gives 
\begin{equation}
	\Pi(\vec{q},\omega_p-i \gamma_p)\approx \Re\Pi(\vec{q},\omega_p)-i\gamma_p \partial_\omega \Re\Pi(\vec{q},\omega)\Big|_{\omega=\omega_p} + i \Im\Pi(\vec{q},\omega_p).
\end{equation}
By substituting this expansion into Eq.(\ref{dielectricRPA}), one finds that the energy of the plasmon is determined from the real part according to
\begin{equation}\label{eq:realpartofplasmonenergyequation}
	1-V(\vec{q})\Re\Pi(\vec{q},\omega_p) = 0,
\end{equation}
whereas the decay rate is a solution of the imaginary part which is given by
\begin{equation}\label{eq:plasmonlifetime}
	\gamma_p = \frac{\Im \Pi(\vec{q},\omega)}{\partial_\omega \Re\Pi(\vec{q},\omega)}\Bigg|_{\omega=\omega_p}.
\end{equation}
\begin{figure}
	\centering
	\includegraphics[width=0.5\textwidth]{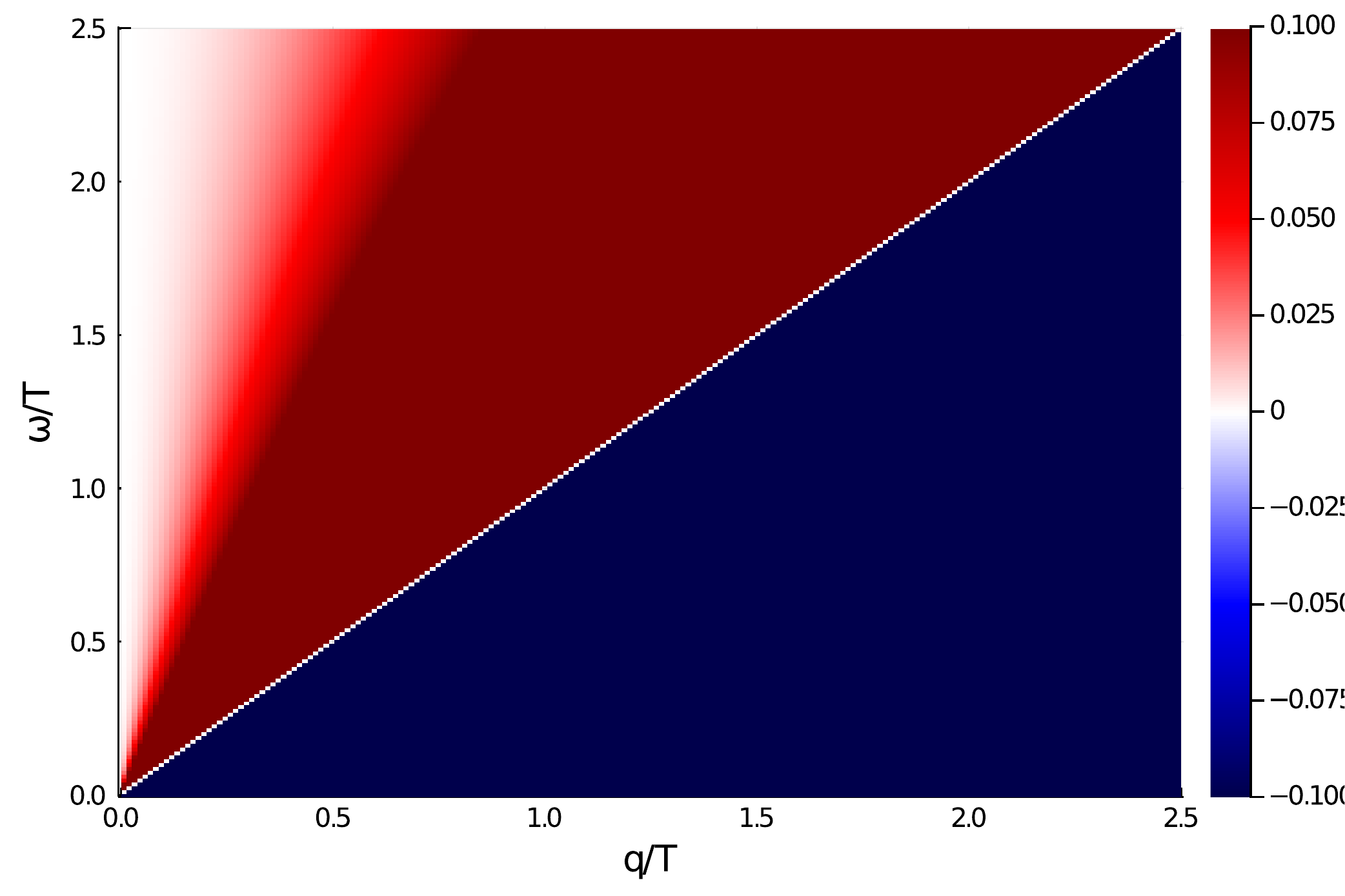}
	\includegraphics[width=0.5\textwidth]{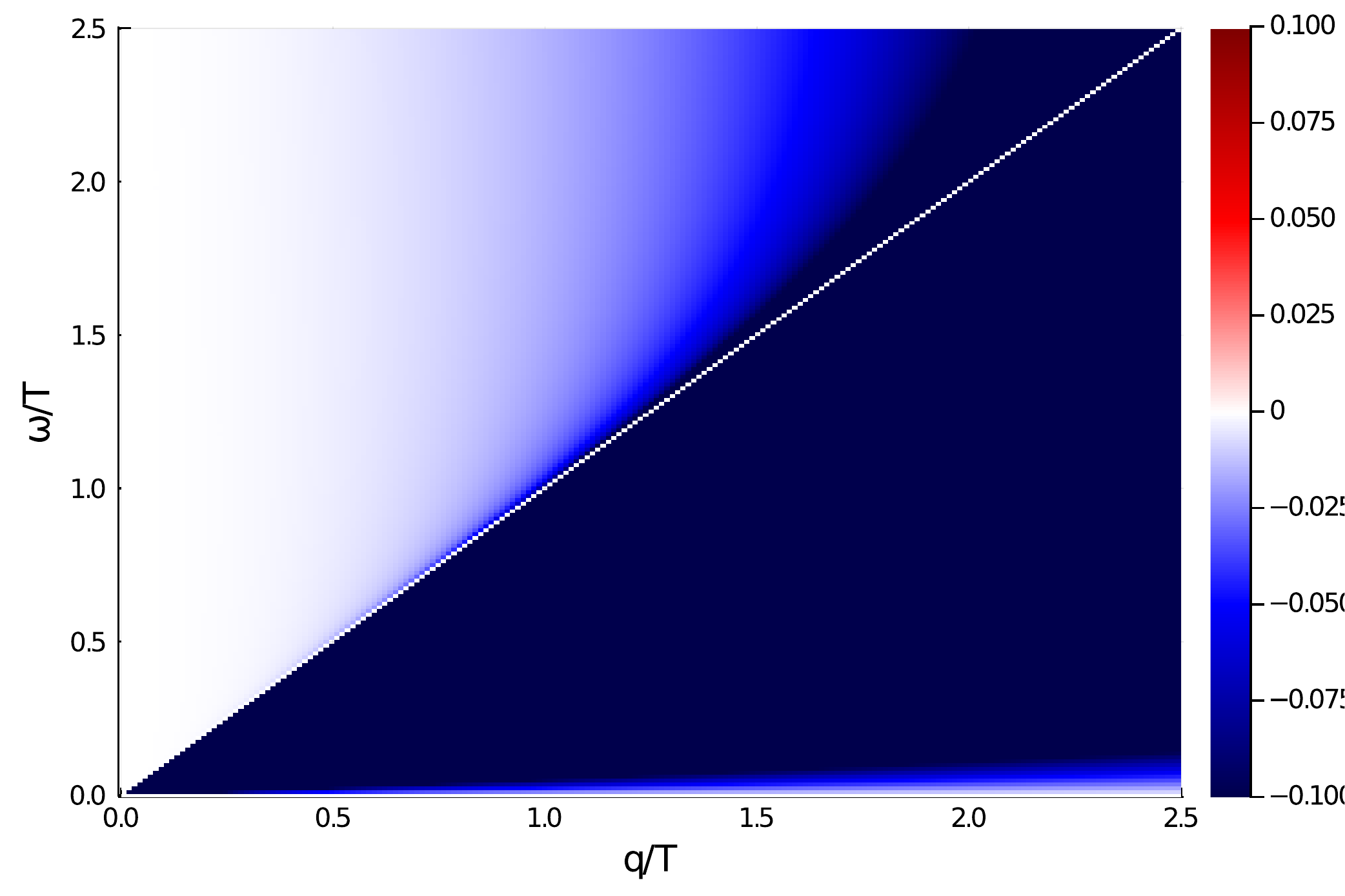}
	\caption{Polarization function, in the unit of temperature $T$, at a non-zero temperature and a chemical potential $\mu/T$ = 1 in the momentum-frequency plane. The above plot shows the real part and the lower plot shows the imaginary part.}
	\label{fig:polarizationfunction}
\end{figure}
A solution to Eq.~\eqref{eq:realpartofplasmonenergyequation} exists only when $\Re \Pi > 0$. In the upper panel of Fig.~\ref{fig:polarizationfunction}, we show the real part of the polarization in the momentum-frequency plane. We observe that $\Re \Pi$ is positive when $\omega>q$. Furthermore, considering Eq.~\eqref{eq:plasmonlifetime}, we find that a stable plasmon solution requires $\Im \Pi = 0$. In the lower panel of  Fig.~\ref{fig:polarizationfunction}, we plot the imaginary part of the polarization function. We observe that although it is not identically zero, it is still negligibly small in the low-momentum limit and $\omega > q$. Consequently, we may expect a long-wavelength underdamped plasmon mode with almost infinitely long lifetime. 

Having made this observation, we  expand the polarization function up to first order in $q/\omega$ by making use of the following expressions:
\begin{eqnarray}
	\mathcal{F}_{\pm\pm}(\vec{p},\vec{q})&=&\frac{1}{2}\left(1+\cos\left(\theta_{\vec{p}+\vec{q}}-\theta_{\vec{q}}\right)\right) \approx 1 ,
	\nonumber\\		
	\mathcal{F}_{\pm\mp}(\vec{p},\vec{q})&=&\frac{1}{2}\left(1-\cos\left(\theta_{\vec{p}+\vec{q}}-\theta_{\vec{q}}\right)\right) \approx \frac{1}{4}\left(\vec{p}\cdot\partial_{\vec{q}}\theta_{\vec{q}}\right)^2,
	\nonumber\\		
	f^0_\lambda(\vec{p}+\vec{q})&\approx& f^0_\lambda(\vec{q})+\vec{p}\cdot\partial_{\vec{q}}f^0_\lambda(\vec{q}),
	\nonumber\\		
	\epsilon_\lambda(\vec{p}+\vec{q})&\approx& \epsilon_\lambda(\vec{q})+\vec{p}\cdot\partial_{\vec{q}}\epsilon_\lambda(\vec{q}).
	\label{eq:longwavelengthexpansion}
\end{eqnarray}
Let us first calculate the real part of the polarization in Eq.~\eqref{eq:polarizationfunction}. The intraband contribution, when $\lambda=\lambda'$, gives
\begin{eqnarray}
	\Re\Pi(\vec{q},\omega) &\approx& N\sum_{\lambda=\pm1} \int\frac{d\vec{p}}{(2\pi)^2} 
	\frac{-\vec{q}\cdot\partial_{\vec{p}}f^0_\lambda(\vec{p})}{\omega}\Big[1+\frac{\vec{q}\cdot\partial_{\vec{p}}\epsilon_\lambda(\vec{p})}{\omega}\Big],\nonumber\\
	&=&\frac{N}{\omega^2} \int\frac{d\vec{p}}{(2\pi)^2} \Big[ 
	f^0_+(\vec{p})(\vec{q}\cdot\partial_{\vec{p}})^2\epsilon_+(\vec{p})-\left(1-f^0_-(\vec{p})\right)(\vec{q}\cdot\partial_{\vec{p}})^2\epsilon_-(\vec{p})\Big], \nonumber\\
	&=& \frac{Nq^2}{\omega^2} \int\frac{pdpd\theta}{(2\pi)^2} \Big[ 
	f^0_+(\vec{p})+\left(1-f^0_-(\vec{p})\right)\Big]\frac{\sin^2\theta}{p}\nonumber\\&=& \frac{Nq^2}{4\pi\omega^2}\int dp\Big[f^0_+(\vec{p})+(1-f^0_-(\vec{p}))\Big]\nonumber\\&=& -\frac{Nq^2}{4\pi\omega^2}T\Big[\text{Li}_1\left(-e^{\mu/T}\right)+\text{Li}_1\left(-e^{-\mu/T}\right)\Big]\nonumber\\&=& \frac{Nq^2}{4\pi\omega^2}T\Big[\log\left(1+e^{\mu/T}\right)+\log\left(1+e^{-\mu/T}\right)\Big]\nonumber\\&=&\frac{Nq^2}{4\pi\omega^2}T\Big[\log\left(2+2\cosh\mu/T\right)\Big]\;.
	\label{eq:realpart}
\end{eqnarray}
In contrast, the interband contribution, when $\lambda=-\lambda'$, gives a logarithmic correction which will be neglected in evaluating the plasmon energy dispersion. For the case of non-zero dopings, at zero temperature, this interband contribution reads
$\frac{Np^2}{16\pi\omega} \log \left(\Big|\frac{\omega-2\mu}{\omega+2\mu}\Big|\right)$.
By substituting Eq.~\eqref{eq:realpart} into Eq.~\eqref{eq:realpartofplasmonenergyequation}, we can derive the dispersion relation for plasmon. It reads 
\begin{equation}
	\omega_p(\vec{q}) = \pm \sqrt{\frac{N}{2}\alpha T q \log\left(2+2\cosh(\mu/T)\right)}.
	\label{eq:plasmonenergy}
\end{equation}
Note that this is the same result as we obtained from the Boltzmann approach in the main text.
\begin{figure}[t]
	\centering
	\includegraphics[width=0.5\textwidth]{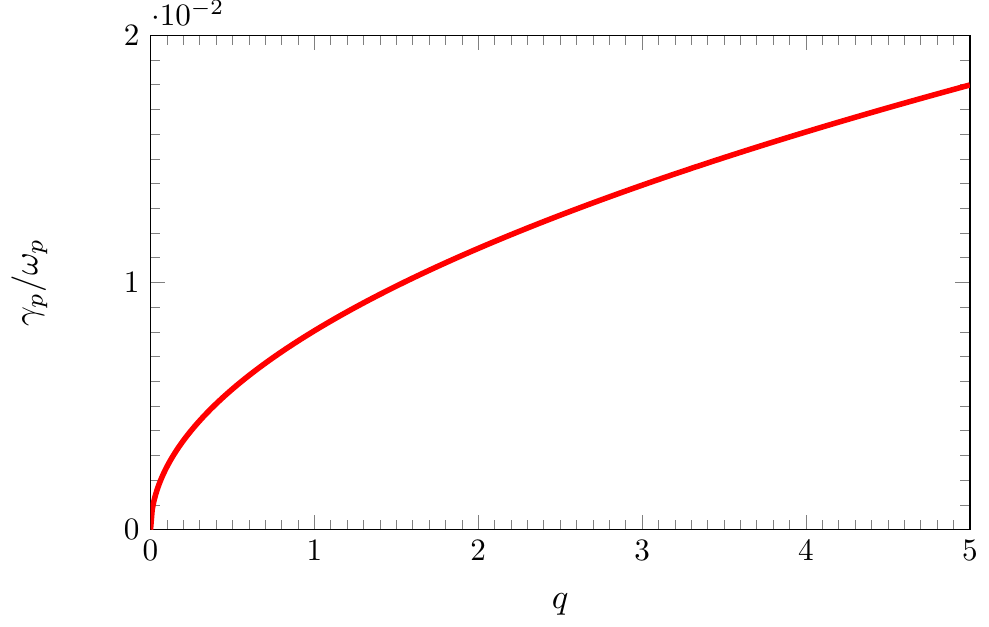}
	\caption{The plot shows the ratio of the plasmon decay rate evaluated in Eq.~\eqref{eq:plasmondecayrate} to the energy dispersion evaluated in Eq.~\eqref{eq:plasmonenergy} for small momenta. It can be observed that, within the low-momentum approximation, the ratio is small. As a result, the plasmon is a well-defined excitation. Here we use $\mu/T = 3$, but it should be noted that this feature is generic for any $\mu/T$.}
	\label{fig:gammatoomega}
\end{figure}

\begin{figure}[t]
	\centering
	\includegraphics[width=0.5\textwidth]{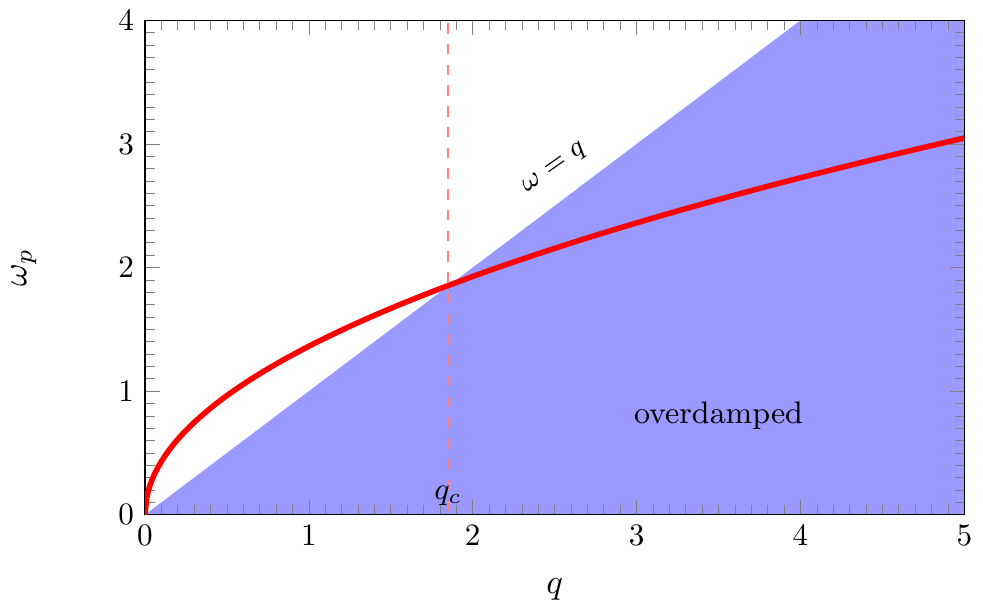}
	\caption{The plot shows the energy dispersion of plasmon. Here we use $\mu/T=3$.}
	\label{fig:plasmonenergy}
\end{figure}

Next, we consider the imaginary part of the polarization function. The main contribution to the imaginary part
is from the interband transition, when $\lambda=-\lambda'$. This gives
\begin{eqnarray}
	\Im\Pi(\vec{q},\omega) &\approx&-N \pi \sum_{\lambda=\pm1} \int\frac{d\vec{p}}{(2\pi)^2} \frac{1}{4}\left(\vec{q}\cdot\partial_{
		\vec{p}}\theta_{\vec{p}}\right)^2 \Big[ 
	f^0_\lambda(\vec{p})-f^0_{-\lambda}(\vec{p}) \Big]  \delta(\omega+\epsilon_\lambda(\vec{p})-\epsilon_{-\lambda}(\vec{p})),\nonumber\\
	&=&-N \pi  \int\frac{d\vec{p}}{(2\pi)^2}\frac{1}{4}\frac{q^2}{p^2}\sin^2(\theta) \Big[\Big(
	f^0_+(\vec{p})-f^0_{-}(\vec{p}) \Big) \delta(\omega+\epsilon_+(\vec{p})-\epsilon_{-}(\vec{p}))+\Big(
	f^0_-(\vec{p})-f^0_{+}(\vec{p}) \Big) \delta(\omega+\epsilon_-(\vec{p})-\epsilon_{+}(\vec{p}))\Big].\nonumber
\end{eqnarray}
Notice that since $f^0_\pm(\vec{p})$ depends only on the magnitude of the momentum $\vec{p}$, let us now denote it by $f^0_\pm(p)$. The angular integral can be performed, giving
\begin{eqnarray}
	\Im\Pi(\vec{q},\omega) &\approx&-\frac{N}{16}q^2 \int_0^{\infty} \frac{dp}{p} \Big[\Big(
	f^0_+(p)-f^0_{-}(p) \Big) \delta(\omega+2p)+\Big(
	f^0_-(p)-f^0_{+}(p) \Big) \delta(\omega-2p)\Big],
	\nonumber\\
	&=&-\frac{N}{16}q^2 \int_0^{\infty} \frac{dp}{p} \Big(
	f^0_+(p)-f^0_{-}(p) \Big) \Big(  \delta(\omega+2p)- \delta(\omega-2p)\Big),\nonumber\\
	&=&-\frac{N}{16}q^2 \int_0^\infty \frac{dp}{p} \Big(
	f^0_+(p)-f^0_{-}(p) \Big) \frac{1}{2} \Big( \delta(p+\omega/2)- \delta(p-\omega/2)\Big),\nonumber\\
	&=& \frac{N}{16}\frac{q^2}{\omega} \left(\frac{1}{e^{\frac{|\omega|/2-\mu}{T}}+1}-\frac{1}{e^{\frac{-|\omega|/2-\mu}{T}}+1}\right)\;.
	\label{eq:imagpol}
\end{eqnarray}
In the limit of zero temperature, this becomes
\begin{equation}
	\Im\Pi^{R}(\vec{q},\omega) \approx -\frac{N}{16}\frac{q^2}{\omega} \Theta(|\omega|-2|\mu|),
\end{equation}
which was found previously in Ref.~\cite{Wunsch2006}. It vanishes when $|\omega|<2|\mu|$, consequently, the long-lived plasmon mode exists in this region.
By substituting the real part in Eq.~\eqref{eq:realpart} and the imaginary part in Eq.~\eqref{eq:imagpol} into Eq.~\eqref{eq:plasmonlifetime}, we find the decay rate of plasmon. It reads
\begin{equation}
	\gamma_{p}(\vec{q}) = -\frac{\pi\omega_p(\vec{q})^2}{8T\log(2+2\cosh\mu/T)}\left(\frac{1}{e^{\frac{|\omega_p(\vec{q})|/2-\mu}{T}}+1}-\frac{1}{e^{\frac{-|\omega_p(\vec{q})|/2-\mu}{T}}+1}\right).
	\label{eq:plasmondecayrate}
\end{equation}

In Fig.~\ref{fig:gammatoomega}, we show the ratio of the decay rate to the energy evaluated above in Eqs.~\eqref{eq:plasmondecayrate}) and~\eqref{eq:plasmonenergy}. We find that $\gamma_p/\omega_p \ll 1$, so, within the low-momentum approximation, the plasmon is a well-defined excitation. However, as we show in Fig.~\ref{fig:plasmonenergy}, the square-root $q$ energy dispersion enters the region $\omega<q$ where the plasmon becomes an over-damped mode. This defined a momentum cutoff above which the plasmon is an unstable mode. The momentum cutoff satisfies
\begin{equation}
	\omega_p(q_c)/q_c = 1\;,
\end{equation}
which translates to
\begin{equation}
	q_c = \frac{N}{2}\alpha T \log(2+2\cosh(\mu/T))\;.
	\label{eq:momentumcutoff}
\end{equation}

\section{Relaxation time of plasmon}
In this section, we are going to derive the relaxation time $\tau_b$ for the plasmon used in the main text. In the companion paper, we derive the Boltzmann-equation for plasmons. It reads
\begin{eqnarray}\label{eq:boltzmannequationboson}
	\partial_{t} b(\vec{x},\vec{p},t) + \vec{v}_b\cdot \partial_{\vec{x}} b(\vec{x},\vec{p},t) &=& -\frac{2\alpha \pi^2 N \omega_p(\vec{p})}{p} \int \frac{d\vec{q}}{(2\pi)^2}  \mathcal{F}_{\lambda\lambda'}(\vec{p}+\vec{q},\vec{q})\delta(\omega_p(\vec{p})+\epsilon_\lambda(\vec{q})-\epsilon_{\lambda'}(\vec{p}+\vec{q})) \nonumber\\&& \Big[f_\lambda(\vec{q})\Big(1-f_{\lambda'}(\vec{p}+\vec{q})\Big)b(\vec{p})  - \Big(1-f_{\lambda}(\vec{q})\Big)f_{\lambda'}(\vec{p}+\vec{q})\Big(1+b(\vec{p})\Big) \Big].\nonumber\\
\end{eqnarray}
By substituting $f=f^0$ and $b = b^0+\delta b$ to the Boltzmann equation above, the left-hand-side becomes
\begin{equation}
	\partial_{t} \delta b(\vec{x},\vec{p},t) + \vec{v}_b\cdot \partial_{\vec{x}} T \partial_T b^0(\vec{p}),
\end{equation}
and the collision integral becomes
\begin{eqnarray}
	-\frac{2\alpha \pi^2 N \omega_p(\vec{p})}{p} \int \frac{d\vec{q}}{(2\pi)^2}  \mathcal{F}_{\lambda\lambda'}(\vec{p}+\vec{q},\vec{q})\delta(\omega_p(\vec{p})+\epsilon_\lambda(\vec{q})-\epsilon_{\lambda'}(\vec{p}+\vec{q}))  \left(f^0_\lambda(\vec{q}) - f^0_{\lambda'}(\vec{p}+\vec{q}) \right) \delta b(\vec{p}) &=& 	 \frac{2\alpha\pi\omega_p(\vec{p})}{p} \Im \Pi \delta b\nonumber\\&=&-2 \gamma_p\delta b,
\end{eqnarray}
where $\gamma_p$ is the decay rate for plasmon in Eq.~\eqref{eq:plasmondecayrate}. To obtain the result above, we approximate the imaginary part of the polarization function at the same level as we did in the previous section. Let us note that in principle there are contributions of the form $-\delta f_+/\tau_{+b}$ and $-\delta f_-/\tau_{-b}$ to the collision integral above. However, an evaluation of the $1/\tau_{\pm b}$ is equivalent to solving the Boltzmann equation. This is beyond the scope of this paper. For this approach, we refer the reader to Ref.~\cite{Kitinan2020}.

\section{Thermo-electric responses of a hybrid system of electrons, holes, and plasmons}
\label{sec:ch1thermoelectricII}
In this section, we discuss the role of the internal electric potential in thermo-electric transport phenomena. One interesting effect of this force-term is the non-local transport response which has been investigated, for example, in Refs.\cite{Mueller2008,KiselevSchmalian2020}. In this paper, we focus on local thermo-electric transport. In this case, the Hartree potential has a direct contribution to thermo-electric transport phenomena via the collective charge-density oscillations or plasmons. Plasmons do not contribute directly to charge transport, because they are electrically neutral. However, they give an extra contribution to the heat current which is given by
\begin{equation}
	\vec{j}^Q_b = \int\frac{d{\vec{p}}}{(2\pi)^2} \vec{v}_b(\vec{p})\omega(\vec{p}) \delta b\;,
\end{equation}
where $\delta b$ is deviation of the distribution function for plasmons from the equilibrium solution.
The deviation $\delta b$ is a solution of the linearized coupled Boltzmann equations. Here $\omega(\vec{p})$ is the energy dispersion of the plasmon given by Eq.~\eqref{eq:plasmonenergy}~\cite{Wunsch2006,Sarma2013}. 

We obtain the coupled system of Boltzmann equations for a hybrid system of the electrons, holes, and plasmons and it takes the form
\begin{eqnarray}
	\left(\partial_t  + \vec{v}_{\lambda}(\vec{p})\cdot\vec{\nabla} \right)f_\lambda(\vec{x},\vec{p},t) &=& \mathcal{C}_\lambda[f_+,f_-,b](\vec{p})\;,\nonumber\\
	\left(\partial_t  + \vec{v}_{b}(\vec{p})\cdot\vec{\nabla}\right)b(\vec{x},\vec{p},t) &=& \mathcal{C}_b[f_+,f_-,b](\vec{p}),\nonumber\\
\end{eqnarray}
where $b(\vec{x},\vec{p},t)$ is the distribution function for the plasmons which the velocity $\vec{v}_b(\vec{p})$. 
To obtain the qualitative feature of the thermo-electric responses, we can resort to the relaxation-time approximation and use
\begin{equation}
	C_\lambda[f_+,f_-,b](\vec{p}) = -\frac{\delta f_{\lambda}}{\tau_{\lambda}} + \frac{\delta f_{-\lambda}}{\tau_{-\lambda}} - \frac{\delta f_{\lambda}}{\tau^{\text{dis}}} + \frac{\delta b}{\tau_b},
\end{equation}
and 
\begin{equation}
	C_b[f_+,f_-,b](\vec{p}) = -\frac{\delta b}{\tau_b/2},
\end{equation}
where $\tau_b$ is the Landau damping of a plasmon into an electron-hole pair. Note that we use the relaxation-time approximation discussed in the main text which respects the required conservation laws.  Within linear-response theory, in the presence of an external electric field $\vec{E}$ and a temperature gradient $\vec{\nabla}T$, we linearize the Boltzmann-equations to linear order in external fields. The deviation of the distribution functions to linear order in the external disturbances can be solved from these Boltzmann-equations. This solution is used to determine the charge and heat currents and, consequently, the thermo-electric coefficients. We find that the expression for the electric conductivity remains the same as in Eq.~\eqref{eq:sigma}, since plasmons do not have a direct contribution to the charge current. However, they can relax electrons, and thus modify the relaxation time for electron-hole drag according to
\begin{eqnarray}
	\frac{1}{\tau_+} &=& \frac{ \frac{1}{\tau'_0}\left( \mathcal{K}_{-}+2\left( \mathcal{T}_-+\mu\mathcal{E}_-\right)\right)}{\left(\mathcal{K}_{-}+2\left( \mathcal{T}_-+\mu\mathcal{E}_-\right)\right)-\left(\mathcal{K}_{+}+2\left( \mathcal{T}_++\mu\mathcal{E}_+\right)\right)},\nonumber\\
	\frac{1}{\tau_-} &=& \frac{-\frac{1}{\tau'_0}\left(\mathcal{K}_{+}+2\left( \mathcal{T}_++\mu\mathcal{E}_+\right)\right)}{\left(\mathcal{K}_{-}+2\left( \mathcal{T}_-+\mu\mathcal{E}_-\right)\right)-\left(\mathcal{K}_{+}+2\left( \mathcal{T}_++\mu\mathcal{E}_+\right)\right)};,\nonumber\\
\end{eqnarray}
where we defined here second-rank tensors
\begin{eqnarray}
	\vec{\vec{\mathcal{K}}}_{+} &=& \int_{\vec{k}} \partial_{\vec{k}} b^0 \vec{v}_{+} \omega({k})\;,\nonumber\\ \vec{\vec{\mathcal{K}}}_{-} &=& \int_{\vec{k}} \partial_{\vec{k}} b^0 \vec{v}_{-} \omega(\vec{k})\;,
\end{eqnarray}
and $\vec{\vec{\mathcal{K}}}_\pm = \mathcal{K}_\pm \vec{\vec{\mathbb{1}}}$.
Note that in calculating $\mathcal{K}_{\pm}$ above, the integrations have the momentum cutoff given by Eq.(\ref{eq:momentumcutoff}). Beyond this value, the quasi-particle become over-damped (see Appendix B) and thus its corresponding distribution function is not well-defined. We further use $\tau'_0 = \tau_b$ since the electron-hole drag results effectively from the Landau damping. 

The themo-electric coefficient has an additional contribution from the drag. It is given by
\begin{equation}
	\alpha = \alpha_{e} + \alpha_{\text{drag}}\;,
\end{equation}
where 
\begin{equation}
	\alpha_{\text{drag}}=\frac{e}{T} \frac{\left(\frac{1}{\tau^{\rm{dis}}}+\frac{2}{\tau_-}\right)\mathcal{K}_{+}+\left(\frac{1}{\tau^{\rm{dis}}}+\frac{2}{\tau_+}\right)\mathcal{K}_{-}}{\frac{2}{\tau^{\rm{dis}}}\left(\frac{1}{\tau^{\rm{dis}}}+\frac{1}{\tau_+}+\frac{1}{\tau_-}\right)}
\end{equation}
and $\alpha_{e}$ is given by Eq.~\eqref{eq:alpha}. We observe that $\alpha_{\text{drag}}=0$ at charge neutrality. Similarly, there are three contributions to the thermal conductivity according to
\begin{equation}
	\kappa =\kappa_{e} + \kappa_{b} + \kappa_{\text{drag}}.
\end{equation}
Here $\kappa_{e}$ is from electrons and holes, which is given by Eq.~\eqref{eq:kappa} and $\kappa_{b}$ is a direct contribution from plasmons, which is given by
\begin{equation}
	\kappa_{b} = -\frac{\tau_b/2}{T}\int_{\vec{k}}   \left(\omega(\vec{k})\right)^2 \vec{v}_{b} \cdot \partial_{\vec{k}} b^0.
\end{equation}
In addition, there is a contribution due to plasmon drag according to
\begin{eqnarray}
	\kappa_{{\rm{drag}}}&=&- \frac{1}{T}\frac{1}{\frac{2}{\tau^{\rm{dis}}}\left(\frac{1}{\tau^{\rm{dis}}}+\frac{1}{\tau_-}+\frac{1}{\tau_+}\right)}\times\nonumber\\&&\int_{\vec{k}} \omega(\vec{k}) \left(\frac{\vec{v}_{-}(\epsilon_--\mu)}{\frac{1}{\tau^{\rm{}dis}}+\frac{2}{\tau_+}}+\frac{\vec{v}_{+}(\epsilon_+-\mu)}{\frac{1}{\tau^{\rm{dis}}}+\frac{2}{\tau_-}}\right) \cdot \partial_{\vec{k}} b^0.\nonumber\\
\end{eqnarray}
Recall that the momentum integrals in $\kappa_{b}$ and $\kappa_{\text{drag}}$ above has a upper bound set by the momentum cutoff given by Eq.(\ref{eq:momentumcutoff}).
Figs.~\ref{fig:transportplasmon} and~\ref{fig:Wiedemann-Franzratio} show the thermo-electric coefficients of the hybrid system of electrons, holes, and plasmons. We observe the enhancement of the Weidemann-Franz ratio not only at the charge-neutrality point but also at high doping due to the extra contribution from the plasmons to the heat transport. 
\end{document}